\documentclass[prd,superscriptaddress,preprintnumbers,amsmath,nofootinbib,preprint]{revtex4}
\usepackage{amsmath, amssymb, graphics,epsfig}

\usepackage{slashed}

\usepackage{hyperref}

\def\bfnabla{\mbox{\boldmath $\nabla$}}

\def\bfsigma{\mbox{\boldmath $\sigma$}}
\def\lQ{\Lambda_{\rm QCD}}
\newcommand{\nn}{\nonumber}
\newcommand{\be}{\begin{equation}}
\newcommand{\ee}{\end{equation}}
\newcommand{\bea}{\begin{eqnarray}}
\newcommand{\eea}{\end{eqnarray}}
\def\al{\alpha}

\def\siml{{\
    \lower-1.2pt\vbox{\hbox{\rlap{$<$}\lower6pt\vbox{\hbox{$\sim$}}}}\ }}

\def\dsl{\,\raise.15ex\hbox{/}\mkern-13.5mu D}

\def\bfnabla{\mbox{\boldmath $\nabla$}}
\def\bfsigma{\mbox{\boldmath $\sigma$}}

\def\lQ{\Lambda_{\rm QCD}}

\newcommand{\eps}{\epsilon}

\newcommand{\cf}{C_F}

\newcommand{\Appendix}[1]%
    {%
     \section{#1}%
      }

\begin{document}

\title{The chromo-polarizabilities of a heavy quark at weak coupling}
\author{Daniel Moreno and Antonio Pineda}
\address{Grup de F\'\i sica Te\`orica, Dept. F\'\i sica and IFAE-BIST, Universitat Aut\`onoma de Barcelona,\\ 
E-08193 Bellaterra (Barcelona), Spain}
\date{\today}

\begin{abstract}
We obtain the renormalization group  improved expressions 
of the Wilson coefficients of the HQET Lagrangian with leading logarithmic approximation to ${\cal O}(1/m^3)$  for the spin-independent sector, which includes the heavy quark chromo-polarizabilites. Our analysis includes the effects induced by spectator quarks. We observe that the numerical impact of these logarithms is very large in most cases. 
\end{abstract}

\maketitle

\section{Introduction}

The expansion in inverse powers of the heavy quark mass is a useful tool for the study of hadrons containing one (or more) heavy quarks. This expansion is formulated more systematically in terms of an effective theory and of its associated effective Lagrangian. For the one-heavy quark sector this effective theory is the Heavy quark effective theory (HQET) \cite{HQET}. Once obtained, its Lagrangian can be applied to physical observables associated to the $B$ or $D$ mesons, such as their spectrum or decays. The HQET Lagrangian is also instrumental in the description of systems with more than one heavy quark, in particular if we fix our attention to the 
heavy quark-antiquark sector (i.e. heavy quarkonium), as the HQET Lagrangian corresponds to one of the building blocks of the Non-relativistic QCD (NRQCD) Lagrangian~\cite{NRQCD,Bodwin:1994jh}. The Wilson coefficients of the HQET Lagrangian operators also enter into the Wilson coefficients of the operators (i.e. the potentials) of the potential NRQCD (pNRQCD) Lagrangian \cite{Pineda:1997bj,Brambilla:1999xf}, 
an effective field theory optimised for the description of heavy quark-antiquark systems near threshold (for reviews see \cite{Brambilla:2004jw,Pineda:2011dg}). Important for us is that the Wilson coefficients we will compute in this paper are necessary ingredients to obtain the pNRQCD Lagrangian with next-to-next-to-next-to-leading log (NNNLL) accuracy, which in turn is the necessary precision to obtain the complete heavy quarkonium spectrum with NNNLL accuracy, and also necessary for the 
computation of the production and annihilation of heavy quarkonium with NNLL precision. Actually, this is one of the main motivations we undertake this work. These results are also instrumental in the determination of higher order logarithms for NRQED bound states, like in 
hydrogen and muonic hydrogen-like atoms.

At present the operator structure of the HQET Lagrangian, and the tree-level values of their Wilson coefficients, is known to ${\cal O}(1/m^3)$ in the case with no massless  quarks \cite{Manohar:1997qy}. The inclusion of light quarks has been considered in \cite{Balzereit:1998am}.
The Wilson coefficients with leading-log (LL) accuracy were computed in \cite{Finkemeier:1996uu,Blok:1996iz,Bauer:1997gs}
to ${\cal O}(1/m^2)$ and in next-to-leading order (NLO) in \cite{Manohar:1997qy} to ${\cal O}(1/m^2)$ (without dimension 6 heavy-light operators). The LL running to ${\cal O}(1/m^3)$ has been considered in Ref. \cite{Balzereit:1998jb,Balzereit:1998am,Balzereit:1998vh}. However, we find several discrepancies. We perform a detailed comparison in Sec. \ref{Sec:Comparison}.

In this paper we obtain the renormalization group (RG) improved expressions 
of the Wilson coefficients of the HQET Lagrangian with LL approximation to ${\cal O}(1/m^3)$  for the spin-independent sector, which includes the heavy quark chromo-polarizabilites. Our analysis includes the effects induced by spectator quarks.

\section{The HQET Lagrangian}

\subsection{HQET Lagrangian without light fermions}
The HQET Lagrangian is defined uniquely up to field redefinitions.
In this paper we use the following HQET Lagrangian density for a quark of mass $ m \gg \lQ$  \cite{Manohar:1997qy}:
\bea
&& 
{\cal L}_{\rm HQET}={\cal L}_g+{\cal L}_{\psi}
\,,
\label{LagHQET}
\\
\nn
\\
&&
{\cal L}_g=-\frac{1}{4}G^{\mu\nu \, a}G_{\mu \nu}^a +
\frac{1}{4}\frac{c_1^{g}}{ m^2} 
g f_{abc} G_{\mu\nu}^a G^{\mu \, b}{}_\alpha G^{\nu\alpha\, c}+
{\cal O}\left(\frac{1}{m^4}\right),
\label{Lg}
\\
\nn
\\
&&
\nn
{\cal L}_{\psi}=
\psi^{\dagger} \Biggl\{ i D_0 + \frac{c_k}{ 2 m} {\bf D}^2 
+ \frac{c_F }{ 2 m} {\bfsigma \cdot g{\bf B}} 
\\
&& \nn
+\frac { c_D}{ 8 m^2} \left({\bf D} \cdot g{\bf E} - g{\bf E} \cdot {\bf D} \right) + i \, \frac{ c_S}{ 8 m^2} 
{\bfsigma \cdot \left({\bf D} \times g{\bf E} -g{\bf E} \times {\bf D}\right) }
\\
&& \nn
+\frac {c_4 }{ 8 m^3} {\bf D}^4 + i c_M\, g { {\bf D\cdot \left[D \times B
\right] + \left[D \times B \right]\cdot D} \over 8 m^3}
+ c_{A_1}\, {g^2}\, {{\bf B}^2-{\bf E}^2 \over 8 m^3}- 
c_{A_2} { {g^2}{\bf E}^2 \over 16 m^3} 
\\ 
&& + c_{W_1}\, g { \left\{ {\bf D^2,\bfsigma
\cdot B }\right\}\over 8 m^3} - c_{W_2}\, g { {\bf D}^i\, {\bf \bfsigma
\cdot B} \, {\bf D}^i  \over 4 m^3}+ c_{p'p}\, g { {\bf \bfsigma \cdot D\, B \cdot D + D
\cdot B\, \bfsigma \cdot D}\over 8 m^3} 
\nn
\\
&&
+ c_{A_3}\, {g^2}\, \frac{1}{N_c}{\rm Tr}\left({{\bf B}^2-{\bf E}^2 \over 8 m^3}\right)- 
c_{A_4}\, {g^2}\, \frac{1}{N_c}{\rm Tr}\left({{\bf E}^2 \over 16 m^3}\right) 
\nn
\\
&&
+ i c_{B_1}\, g^2\, { {\bf \bfsigma \cdot \left(B
\times B - E \times E \right)} \over 8 m^3}
- i c_{B_2}\, g^2\, { {\bf \bfsigma \cdot \left( E \times E \right)} \over 8 m^3}
\Biggr\} \psi+
{\cal O}\left(\frac{1}{m^4}\right)
\,.
\eea
Here $\psi$ is the NR fermion field represented by a Pauli spinor. The components of the vector $\bfsigma$ are the Pauli matrices. We define
$i D^0=i\partial^0 -gA^0$, $i{\bf D}=i\bfnabla+g{\bf A}$,
${\bf E}^i = G^{i0}$ and ${\bf B}^i = -\eps_{ijk}G^{jk}/2$, where $\eps_{ijk}$ is
the three-dimensional totally antisymmetric tensor\footnote{
In dimensional regularization several prescriptions are possible for the $\eps_{ijk}$ tensors and $\bfsigma$, and the same prescription as for the calculation of the Wilson coefficients must be used.}
with $\eps_{123}=1$ and $({\bf a} \times {\bf b})^i \equiv \eps_{ijk} {\bf a}^j {\bf b}^k$. Note also that we have rescaled by a factor $1/N_c$ the coefficients $c_{A_{3,4}}$, compared with the definitions in \cite{Manohar:1997qy}. In general, we will refer to the $c_{A_i}$ as the chromo-polarizabilities.

\subsection{HQET Lagrangian with massless fermions}
We now include $n_f$ massless fermions to the HQET Lagrangian. The Lagrangian now has the following structure:
\bea
&& 
{\cal L}_{\rm HQET}={\cal L}_g+{\cal L}_{\psi}
+{\cal L}_l,
\label{LagHQETnf}
\\
&&
{\cal L}_l = \sum_{i=1}^{n_f} \bar q_i i \dsl q_i 
+\frac{\delta {\cal L}^{(2)}_l}{m^2}
+\frac{\delta {\cal L}^{(2)}_{\psi l}}{m^2}
+\frac{\delta {\cal L}^{(3)}_{ l}}{m^3}
+\frac{\delta {\cal L}^{(3)}_{\psi l}}{m^3} +
{\cal O}\left(\frac{1}{m^4}\right)
.
\eea
The complete set of operators at ${\cal O}(1/m^2)$ can be found in \cite{Bauer:1997gs}. They read
\bea
&& 
\delta {\cal L}^{(2)}_{\psi l}
=\frac{c_1^{hl} }{8}\, g^2 \,\sum_{i=1}^{n_f}\psi^{\dagger} T^a \psi  \
\bar{q}_i\gamma_0 T^a q_i 
+\frac{c_2^{hl} }{ 8}\, g^2 \,\sum_{i=1}^{n_f}\psi^{\dagger}\gamma^\mu\gamma_5
T^a \psi_1 \ \bar{q}_i\gamma_\mu\gamma_5 T^a q_i 
\nn
\\
&& \qquad\qquad
+\frac{c_3^{hl}}{ 8}\, g^2 \,\sum_{i=1}^{n_f}\psi^{\dagger} \psi \ \bar{q}_i\gamma_0 q_i
+\frac{c_4^{hl}}{ 8}\, g^2 \,\sum_{i=1}^{n_f}\psi^{\dagger}\gamma^\mu\gamma_5
\psi \ \bar{q}_i\gamma_\mu\gamma_5 q_i,
\label{Lhl}
\\
&&
\delta {\cal L}^{(2)}_l=\frac{c_D^{l}}{4} \bar q_i \gamma_{\nu} D_{\mu} G^{\mu \nu}
 q_i 
\nn
\\
&& \qquad\qquad
+\frac {c_1^{ll}}{ 8} \, g^2 \,
\sum_{i,j =1}^{n_f} \bar{q_i} T^a \gamma^\mu q_i \ \bar{q}_j T^a \gamma_\mu q_j  
+\frac{c_2^{ll}}{ 8} \, g^2 \,
\sum_{i,j=1}^{n_f}\bar{q_i} T^a \gamma^\mu \gamma_5 q_i \ \bar{q}_j T^a \gamma_\mu \gamma_5 q_j 
\nn
\\ 
&& \qquad\qquad
+ \frac{c_3^{ll} }{ 8}\, g^2 \,
\sum_{i,j=1}^{n_f} \bar{q_i}  \gamma^\mu q_i \ \bar{q}_j \gamma_\mu q_j 
+ \frac{c_4^{ll}}{ 8}  \, g^2 \,
\sum_{i,j=1}^{n_f}\bar{q_i} \gamma^\mu \gamma_5  q_i \ \bar{q}_j \gamma_\mu \gamma_5 q_j.
\label{Ll}
\eea

As we will discuss later, $c_D^l$ and the light-light operators would contribute at NLL. Therefore, we will 
not consider them any further. We only have to discuss the $1/m^3$ heavy-light operators 
(i.e. those of dimension 7). These have been studied in \cite{Balzereit:1998am}. We do not try to make an exhaustive analysis of all of them. We will only consider in detail those that have LL running and affect the running of the chromo-polarizabilities. 
 
The operators relevant for this calculation can be found in Eq. (10) of \cite{Balzereit:1998am}. After disregarding some of 
them because of its spin dependence or because they are proportional to the energy of the heavy quark (so they become subleading after using the equations of motion), we find that, in QCD, the only relevant operators are:
\begin{equation}
 \mathcal{M}_{4\pm}^{(3h)s/o}=\pm g^2[\bar q\gamma^\mu \mathcal{C}_{s/o}^a q][\bar h_v  \mathcal{C}_{s/o}^a i D_\mu^{\pm} h_v]
\,,
\end{equation}

\begin{equation}
 \mathcal{M}_{6\pm}^{(3h)s/o}=\pm g^2[\bar q i\sigma^{\mu\lambda}v_\lambda\mathcal{C}_{s/o}^a q][\bar h_v  \mathcal{C}_{s/o}^a i D_\mu^{\pm} h_v]
\,,\end{equation}

\begin{equation}
 \mathcal{M}_{2\pm}^{(3l)s/o}=\pm g^2[\bar q \mathcal{C}_{s/o}^a (iv D^{\pm}) q][\bar h_v  \mathcal{C}_{s/o}^a h_v]
\,,\end{equation}

\begin{equation}
 \mathcal{M}_{3\pm}^{(3l)s/o}=\pm g^2[\bar q \slashed v \mathcal{C}_{s/o}^a (iv D^{\pm}) q][\bar h_v  \mathcal{C}_{s/o}^a h_v]
\,,\end{equation}

\begin{equation}
 \mathcal{M}_{4\pm}^{(3l)s/o}=\pm g^2[\bar q i\sigma^{\lambda\nu}v_\lambda \mathcal{C}_{s/o}^a iD^{\pm}_\nu q][\bar h_v  \mathcal{C}_{s/o}^a h_v]
\,,\end{equation}
where $iD_\mu^+=i \stackrel{\rightarrow}{\partial}_\mu - g A_\mu^a T^a$ and $iD_\mu^-=i\stackrel{\leftarrow}{\partial}_\mu +gA_\mu^a T^a$. The arrows over the derivatives indicate that the covariant derivatives act over fields in the left/right hand depending on the direction of the arrow (they only act over heavy quark fields
or over light quark fields), $\mathcal{C}_s^a=1$ and $\mathcal{C}_o^a=T^a$ and $\sigma^{\mu\nu}=\frac{i}{2}[\gamma^\mu,\gamma^\nu]$. In our case, we work in the 
rest frame, so that $v^\mu=(1,\bf 0)$ and $h_v \equiv \psi$. 
It is also understood that in the octet case the covariant derivative stands left/right of the color matrix when acting to the left/right. Moreover, we are in the heavy-quark sector, and not in the antiquark one, so we can project to this sector. In practice 
this is equivalent to take the $\gamma^0$ which appears from $\bar h_v\equiv \bar \psi$ equal to one, $\gamma^0 \rightarrow 1$. After all these simplifications, the previous operators can be written as:

\begin{equation}
 \mathcal{M}_{4\pm}^{(3h)s/o}=\pm g^2[\bar q\gamma^\mu \mathcal{C}_{s/o}^a q][\psi^\dagger \mathcal{C}_{s/o}^a i D_\mu^{\pm} \psi]
\,,\end{equation}

\begin{equation}
 \mathcal{M}_{6\pm}^{(3h)s/o}=\pm ig^2[\bar q \sigma^{i 0} \mathcal{C}_{s/o}^a q][\psi^\dagger \mathcal{C}_{s/o}^a i D_i^{\pm} \psi]
\,,\end{equation}

\begin{equation}
 \mathcal{M}_{2\pm}^{(3l)s/o}=\pm g^2[\bar q \mathcal{C}_{s/o}^a (iD_0^{\pm}) q][\psi^\dagger  \mathcal{C}_{s/o}^a \psi]
\,,\end{equation}

\begin{equation}
 \mathcal{M}_{3\pm}^{(3l)s/o}=\pm g^2[\bar q \gamma^0 \mathcal{C}_{s/o}^a (i D_0^{\pm}) q][\psi^\dagger  \mathcal{C}_{s/o}^a \psi]
\,,\end{equation}

\begin{equation}
 \mathcal{M}_{4\pm}^{(3l)s/o}=\pm ig^2[\bar q \sigma^{0 i}\mathcal{C}_{s/o}^a iD^{\pm}_i q][\psi^\dagger\mathcal{C}_{s/o}^a \psi]
\,.
\end{equation}
We then have
\be
\delta {\cal L}^{(3)}_{\psi l}=\sum_i d_i^{hl}O_i
\,,\ee
where the $O_i$ are linear combinations of all possible $ \mathcal{M}$ to be defined later. 
We are not interested in all of them, but only those that may get LL running and contribute to the chromo-polarizabilities. We discuss this issue in Sec.~\ref{Sec:LLlight}. 

\section{Compton scattering}
\label{Sec:Compton}

In order to prepare the computation of the anomalous dimensions of the heavy quark chromo-polarizabilities, it is useful to consider the Compton scattering of a heavy quark with a gluon. In this section we study the spin-independent part of the Compton effect in QCD, which is the scattering of a gluon with a heavy quark, $Qg\rightarrow Qg$. We will 
compute it at tree level up to $\mathcal{O}(1/m^3)$ in the mass expansion and in the Coulomb gauge (though obviously the Compton scattering is a gauge independent object). 
We will consider the incoming and 
outgoing quarks with four-momentum $p=(E_1,\bf p)$ and $p'=(E_1',\bf p\,')$. Gluon four-momenta will be considered as being outgoing and labeled 
as $k_1$, $i$, $a$ and $k_2$, $j$, $b$ with respect to color and vector indices. This also implies the on-shell condition 
$k_1^0= -|\bf k_1|$ and $k_2^0= |\bf k_2|$. We work in the incoming quark rest frame, i.e $E_1=0$ and $\bf p=0$, so $\bf p\,'=-(\bf k_1 + \bf k_2)$ 
and $E_1'=-(k_1^0 + k_2^0)$. In addition, let's define the unit vectors $\bf n_1=\bf k_1/|\bf k_1|$ and 
$\bf n_2=\bf k_2/|\bf k_2|$. The relation
\begin{equation}
 |{\bf k}_2|=\frac{|{\bf k}_1|}{1+\frac{|{\bf k}_1|}{m}(1 + {\bf n}_1\cdot{\bf n}_2)}
 = |{\bf k}_1|\left(1-\frac{|{\bf k}_1|}{m}(1 + {\bf n}_1\cdot{\bf n}_2)\right)
\end{equation}
also holds from four-momenta conservation.

\begin{figure}[!htb]
	\includegraphics[width=0.80\textwidth]{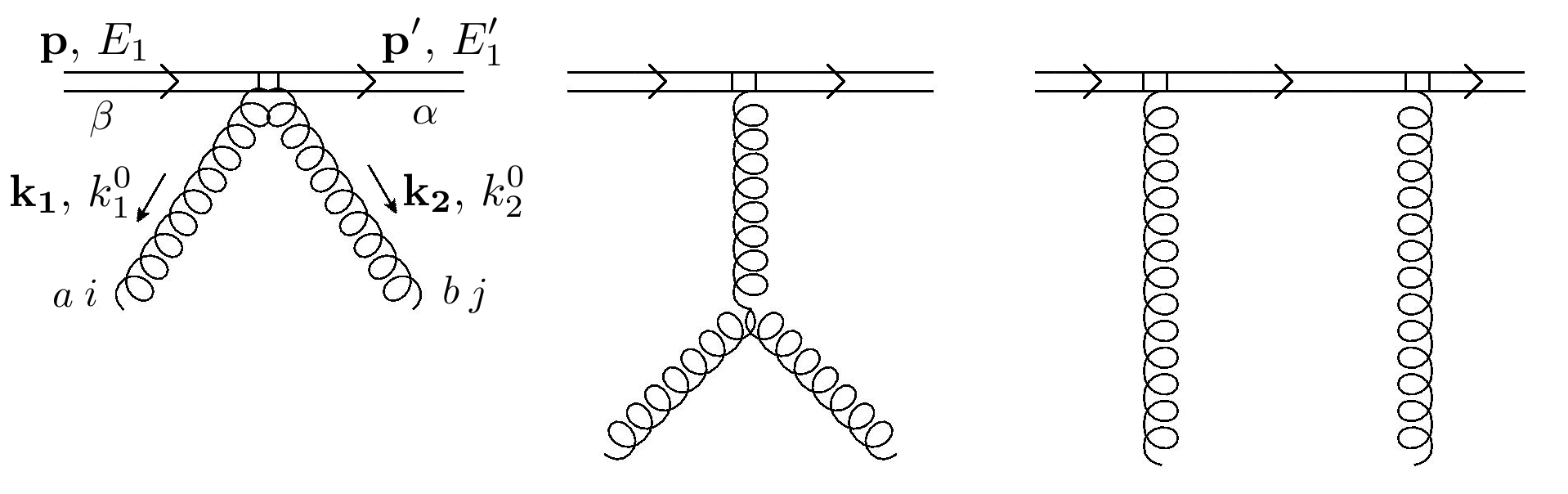} 
	%
	%
	%
\caption{Topologies of the tree level diagrams contributing to the Compton scattering to ${\cal O}(1/m^3)$. Diagrams are generated from these topologies by considering all possible vertex and 
kinetic insertions contributing to ${\cal O}(1/m^3)$.
\label{PlotsCompton}}   
\end{figure}

With appropriate Wilson coefficients (to ${\cal O}(1/m^3)$) the topologies of the diagrams we have to consider for such computation can be found in Fig.~\ref{PlotsCompton}.
Overall, we obtain 
\bea
{\cal A}^{ij\,ab}&=&-ig^2\delta^{ij}\frac{1}{|{\bf k}_1|}\frac{1}{1+{\bf n}_1\cdot{\bf n}_2}[T^a,T^b]_{\alpha\beta}
\nn
\\
&&
-c_k\frac{ig^2}{2m}\delta^{ij}\{T^a,T^b\}_{\alpha\beta} -\frac{ig^2}{2m}\delta^{ij} [T^a,T^b]_{\alpha\beta}
\nn
\\
&&
+ \frac{ig^2}{4m^2}|{\bf k}_1|(\delta^{ij}(2-c_F^{2}{\bf n}_1\cdot{\bf n}_2) + c_F^{2}{\bf n}_2^i {\bf n}_1^j)[T^a, T^b]_{\alpha\beta}
\nn
\\
&&
+c_1^g \frac{3ig^2}{2m^2}[T^a,T^b]_{\alpha\beta}|{\bf k}_1|
\left( \delta^{ij} - \frac{{\bf n}_2^i {\bf n}_1^j}{1+{\bf n}_1\cdot{\bf n}_2}\right)
\nn
\\
&&
+ \frac{ig^2}{16m^3}|{\bf k}_1|^2(\delta^{ij}( (4c_4 + 4c_M -2c_{A_1} -c_{A2} + 2c_Sc_F)
\nn
\\
&&
+  (4c_4-2c_{A_1}){\bf n}_1\cdot{\bf n}_2) 
+ 2c_{A_1}{\bf n}_2^i {\bf n}_1^j)\{T^a,T^b\}_{\alpha\beta}
\nn
\\
&&
 -c_F^{\,2}\frac{ig^2}{8m^3}|{\bf k}_1|^2(\delta^{ij}{\bf n}_1\cdot{\bf n}_2 - {\bf n}_2^i {\bf n}_1^j)({\bf n}_1\cdot{\bf n}_2)\{T^a, T^b\}_{\alpha\beta}
\nn
\\ 
&&
- \frac{ig^2}{8m^3}|{\bf k}_1|^2(\delta^{ij}(2 - c_F^{2}{\bf n}_1\cdot{\bf n}_2) + c_F^{2} {\bf n}_2^i {\bf n}_1^j)(1 + {\bf n}_1\cdot{\bf n}_2) [T^a,T^b]_{\alpha\beta}
\nn
\\
&&- c_1^g \frac{3ig^2}{4m^3}[T^a,T^b]_{\alpha\beta}|{\bf k}_1|^2
( \delta^{ij}(1 + {\bf n}_1\cdot{\bf n}_2) -{\bf n}_2^i {\bf n}_1^j)
\,.
\eea

We first observe that $c_D$ does not appear explicitly in the computation. It only appears implicitly through $c_M$ (as $c_M$ is related 
with $c_D$ by reparameterization invariance). From the above result we also observe that $c_{A_2}$ and $c_M$ always appear in the same combination: $\bar c_{A_2}\equiv c_{A_2}-4c_M$, in observables. Since $c_M$ 
is gauge dependent, only $\bar c_{A_2}$ can be considered physical. Actually, for the observables we have studied we see that 
$c_{A_2}$ always appear in the combination $\tilde c_{A_2}\equiv 2c_{A_1}+\bar c_{A_2}$. This happens for the Compton scattering but also for heavy quarkonium. A similar analysis for the 
elastic scattering of a light and heavy quark shows that $\bar c_1^{hl}\equiv c_D+c_1^{hl}$ is physical. Note, however, that $c_D$ and $c_1^{hl}$ individually are not, indeed they are gauge
dependent (for instance see \cite{Pineda:2001ra} for a discussion on this issue).

For QED we obtain
\bea
 {\cal A}^{ij\,ab}&=&-c_k\frac{ig^2}{m}\delta^{ij}
\nn
\\
&&
+ \frac{ig^2}{8m^3}|{\bf k}_1|^2\delta^{ij}( (2 + 4c_4  -2c_{A_1} +2c_Sc_F + 4c_M -c_{A_2} - 2c_F^{2})
\nn
\\
&&+ (2 + 4c_4^{(1)}  -2c_{A_1} + 2c_F^{2}c_k - 2c_F^{2}){\bf n}_1\cdot{\bf n}_2)
\nn
\\
&&
 +\frac{ig^2}{4m^3}|{\bf k}_1|^2 \delta^{ij}(c_F^{2}(1-{\bf n}_1\cdot{\bf n}_2)-1)(1 + {\bf n}_1\cdot{\bf n}_2)
\nn
\\
&&
 + \frac{ig^2}{4m^3}|{\bf k}_1|^2  ( (c_F^{2} + c_{A_1} - c_F^{2}c_k) + c_F^{2}{\bf n}_1\cdot{\bf n}_2){\bf n}_2^i {\bf n}_1^j
 \eea
Note that there is no ${\cal O}(1/m^0,1/m^2)$ contribution. Setting the Wilson coefficients to their tree level values we obtain
\bea
{\cal A}^{ij\,ab}&=&-\frac{ig^2}{m}\delta^{ij}
\nn
\\
&&+ \frac{ig^2}{2m^3}|{\bf k}_1|^2\delta^{ij}(1  + {\bf n}_1\cdot{\bf n}_2)
\nn
\\
&&
-\frac{ig^2}{4m^3}|{\bf k}_1|^2(\delta^{ij}{\bf n}_1\cdot{\bf n}_2 - {\bf n}_2^i {\bf n}_1^j)(1 + {\bf n}_1\cdot{\bf n}_2)
\,.
\eea
This expression agrees with Eq. (19) in \cite{Balk:1993ev}.

For completeness, we also define polarizabilities. The concept of polarizability is potentially ambiguous, as it is defined after subtracting what are called Born terms (which indeed can be defined in several ways) from the Compton scattering computation. Indeed this discussion already appears in the context of QED and the elastic scattering of photons with protons (see for instance \cite{Scherer:1996ux,Fearing:1996gs,Nevado:2007dd}). 
One possible definition is the one used in \cite{Hill:2012rh}, which adapted to the notation of our paper reads
\begin{equation}
\frac{4m^3}{\alpha}\alpha_{E1}\equiv -c_{A_1} - \frac{\bar c_{A_2}}{2} + c_F^2-c_F+1
\,,
\end{equation}
\begin{equation}
\frac{4m^3}{\alpha}\beta_{M1}\equiv -1  + c_{A_1} 
\,.
\end{equation} 

Note that, even in QED, the polarizabilities are not low energy constants, as they depend on the renormalization scale. 

The above analysis gives us the set of Wilson coefficients and its combinations that appear in physical observables. Those are the ones for which we will compute the anomalous dimensions:
$\{c_{A_1},\bar c_{A_2},c_{A_3},c_{A_4},c_4\}$.

\section{Computation of the $1/m^3$ anomalous dimension}
\label{Sec:RGE}
We want to determine the anomalous dimension of the  $1/m^3$ Wilson coefficient operators with ${\cal O}(\al)$ 
accuracy. 

In principle, one would like to only compute irreducible diagrams. Nevertheless, in such situation we would need to consider a more extense basis of operators, including those that vanish on-shell. Therefore, we will then instead also consider reducible diagrams in a computation that resembles the one of a S-matrix element. In our case we will compute the divergent part of the elastic scattering of the heavy quark with a tranverse gluon at one-loop. These divergences cancel with divergences of the Wilson coefficients determining the anomalous dimension. The computation is organized according to the powers of $1/m$ up to ${\cal O}(1/m^3)$ by considering all possible insertions of the HQET Lagrangian operators. This statement requires some qualifications once $1/m^3$ operators with light fermions fields are involved. For those we do not seek the anomalous dimension of all their Wilson coefficients. To start with we do not consider spin-dependent ones but even for the spin-independent operators we are not exhaustive in the search of a complete basis of operators. The reason is that most of them start to contribute at NLL, playing a subleading role in heavy quarkonium physics. We will only consider those that contribute at LL to the polarizabilities. 

In order to check some parts of the computation we will also compute the elastic scattering of the heavy quark with a longitudinal gluon. This allows to check the combinations $2c_{A_1}+c_{A_2}$ and $2c_{A_3}+c_{A_4}$.

Please note that the matching coefficients of the
kinetic term are protected by reparameterization invariance
($c_k=c_4=1$ to any order in perturbation theory) \cite{Luke:1992cs}. Nevertheless, we will often keep them explicit for tracking purposes. We will also compute the running 
of $c_4$ explictely at one loop as a check. 

In principle, $c_M$ is also fixed by reparameterization invariance. It was originally determined in \cite{Manohar:1997qy}. Recently, a new result, 
$2c_M=c_D-c_F$, 
was obtained \cite{Hill:2012rh}, which differs by a sign of the old one. As we have already discussed, the computation of a physical observable, like the Compton scattering, does not allow to determine the RG equation of $c_M$, as it always appears in the combination $\bar c_{A_2}=c_{A_2}-4c_M$. Working off-shell, our computation in QED in the next section indeed confirms (at one loop and in the Coulomb gauge) $2c_M=c_D-c_F$. Nevertheless, we cannot perform a similar check in QCD. The constraints imposed by reparameterization invariance once operators with light fermion fields are included have also been studied in \cite{Hill:2012rh} for the QED case where the relation $d_1^{hl}=c_1^{hl}/16$ (see Eq. (\ref{O1})) was deduced. Nevertheless, such Wilson coefficient does not enter into the running 
of the chromo-polarizabilities. 

We will perform the computation in the Coulomb gauge. On the one hand this will significantly reduce the number of diagrams to compute. On the other hand the complexity of each of them increases. It also makes difficult to use standard routines for computations of diagrams designed for Feynman gauges and more relativistic-like setups. In the Coulomb gauge the normalization of the quarks and gluon fields and $g$ read (in this paper we define 
$D=4+2\epsilon$)
\bea
\nn
Z_{A^0}^{-1/2}=Z_g&=&1+\frac{11}{6}C_A\frac{\al}{4\pi}\frac{1}{\epsilon} -\frac{2}{3}T_Fn_f\frac{\al}{4\pi}\frac{1}{\epsilon} \,,
\quad Z_{\bf A}^{\frac{1}{2}}=1-\frac{C_A}{2}\frac{\al}{4\pi}\frac{1}{\epsilon}-\frac{2}{3}T_Fn_f\frac{\al}{4\pi}\frac{1}{\epsilon} 
\,,
\\
Z^2_gZ_{\bf A}&=&1+\frac{8}{3}C_A\frac{\al}{4\pi}\frac{1}{\epsilon}
\,,
\quad 
Z_l=1+C_F\frac{\al}{4\pi}\frac{1}{\epsilon}
\,, 
\quad
Z_h=1+\frac{{\bf p}^2}{m^2}\frac{4}{3}C_F\frac{\al}{4\pi}\frac{1}{\epsilon}
\,,
\eea
where
\begin{eqnarray}
  \cf=\frac{N_c^2-1}{2N_c}=\frac{4}{3}\,, \qquad C_A = N_c=3\,.
\end{eqnarray}

\begin{figure}[!htb]
	\includegraphics[width=1.00\textwidth]{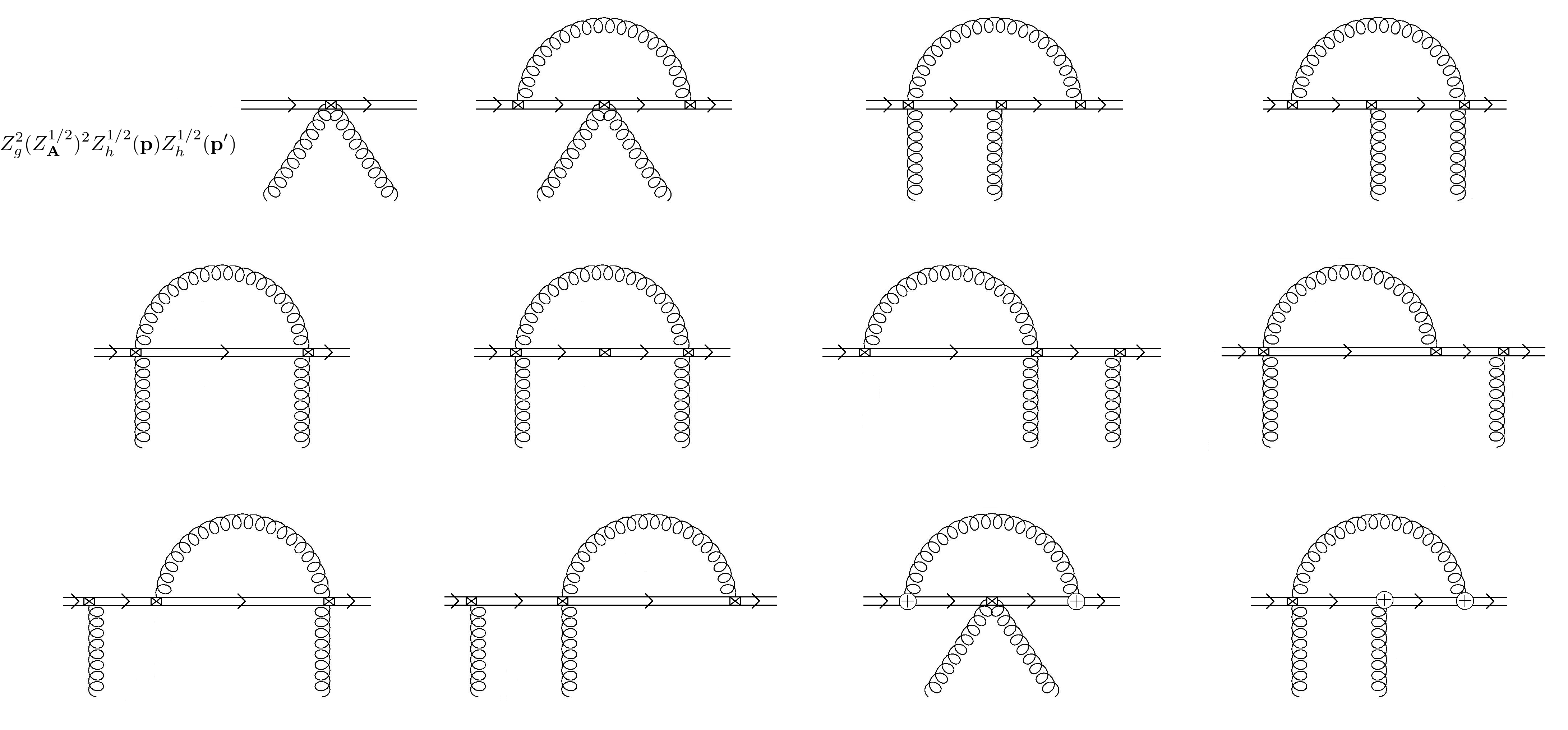}
	\includegraphics[width=1.00\textwidth]{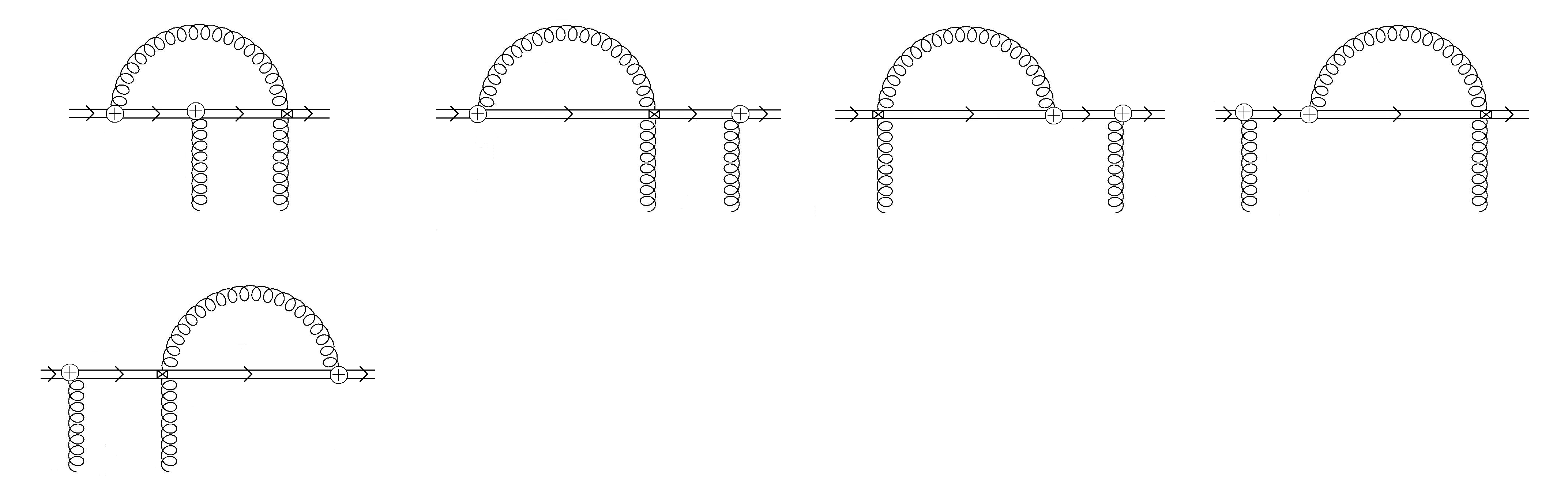}
	%
	%
	%
	%
	%
\caption{The first diagram is the tree-level diagram multiplied by the prefactor 
$Z_g^2 Z_{\bf A}Z_h^{1/2}({\bf p})Z_h^{1/2}({\bf p}')$. 
The rest are the one loop diagrams contributing to the anomalous dimensions of the $1/m^3$ Wilson coefficients in QED. Circles with crosses represent $c_F$ vertices, the other vertices are 
proportional to $c_k$. Crosses in the propagators represent kinetic insertions.  Double-line propagators represent the heavy fermion. The wavy-line stands for transverse photons only in this figure. Note that in some cases one has to Taylor expand in the energy.
\label{PlotsQED}}   
\end{figure}

\subsection{QED case}
As a warm-up, we first consider the pure QED case ($C_F=1$, $C_A=0$ and $n_f=0$). The diagrams that contribute can be found in Fig.~\ref{PlotsQED}.

For the reducible diagrams we have to keep in mind that the sub-irreducible part of the diagram can be Taylor expanded in powers of the energy. When computing, as expected, we find that the non-local terms are finite and all divergences can be absorbed by local counterterms that correspond to operators of the Lagrangian\footnote{If we only compute irreducible diagrams, we would need a larger number of operators, in particular those that vanish on-shell.}.
The result is the following
\bea
c_{M,B}&=&c_{M}+c_k^3\frac{2}{3}\frac{\al}{\pi}\frac{1}{\epsilon}
\,,
\qquad
c_{A_1,B}=c_{A_1}-c_k^3\frac{8}{3}\frac{\al}{\pi}\frac{1}{\epsilon}
\,,
\qquad
c_{A_2,B}=c_{A_2}+c_k^3\frac{40}{3}\frac{\al}{\pi}\frac{1}{\epsilon}
\,,
\eea
where the subscript $B$ stands for the bare Wilson coefficient, whereas the renormalized 
Wilson coefficients do not have an associated subscript. 

$c_4$ does not renormalize. This we explicitly check. Overall, we obtain the following RG equations: 
\bea
\nu\frac{d}{d\nu}c_{M}
=
-c_k^3\frac{4}{3}\frac{\al}{\pi}
\,,
\qquad
\nu\frac{d}{d\nu}c_{A_1}
=
+c_k^3\frac{16}{3}\frac{\al}{\pi}
\,,
\qquad
\nu\frac{d}{d\nu}c_{A_2}
=
-c_k^3\frac{80}{3}\frac{\al}{\pi}
\,.
\eea

Note that in order to determine the running of $c_M$ and $c_{A_2}$ separately, we had to consider the gluons to be off-shell, otherwise we can not distinguish the Feynman rule of $c_M$ and $c_{A_2}$ but only of the physical quantity $\bar c_{A_2}$. In QED this seems to be fine. For instance the running of $c_D$ in QED happens to be equal in the Coulomb and in the Feynman gauge (see the discussion in
 \cite{Pineda:2001ra}). For QCD working off-shell produces divergences that violate reparameterization invariance for $c_M$. Therefore, in the following we 
will restrict ourselves to the analysis of physical quantities, i.e. $\bar c_{A_2}$.

\subsection{$1/m^3$ light fermions: LL running of $d_i^{hl}$}
\label{Sec:LLlight}

\begin{figure}[!htb]
	\includegraphics[width=1.00\textwidth]{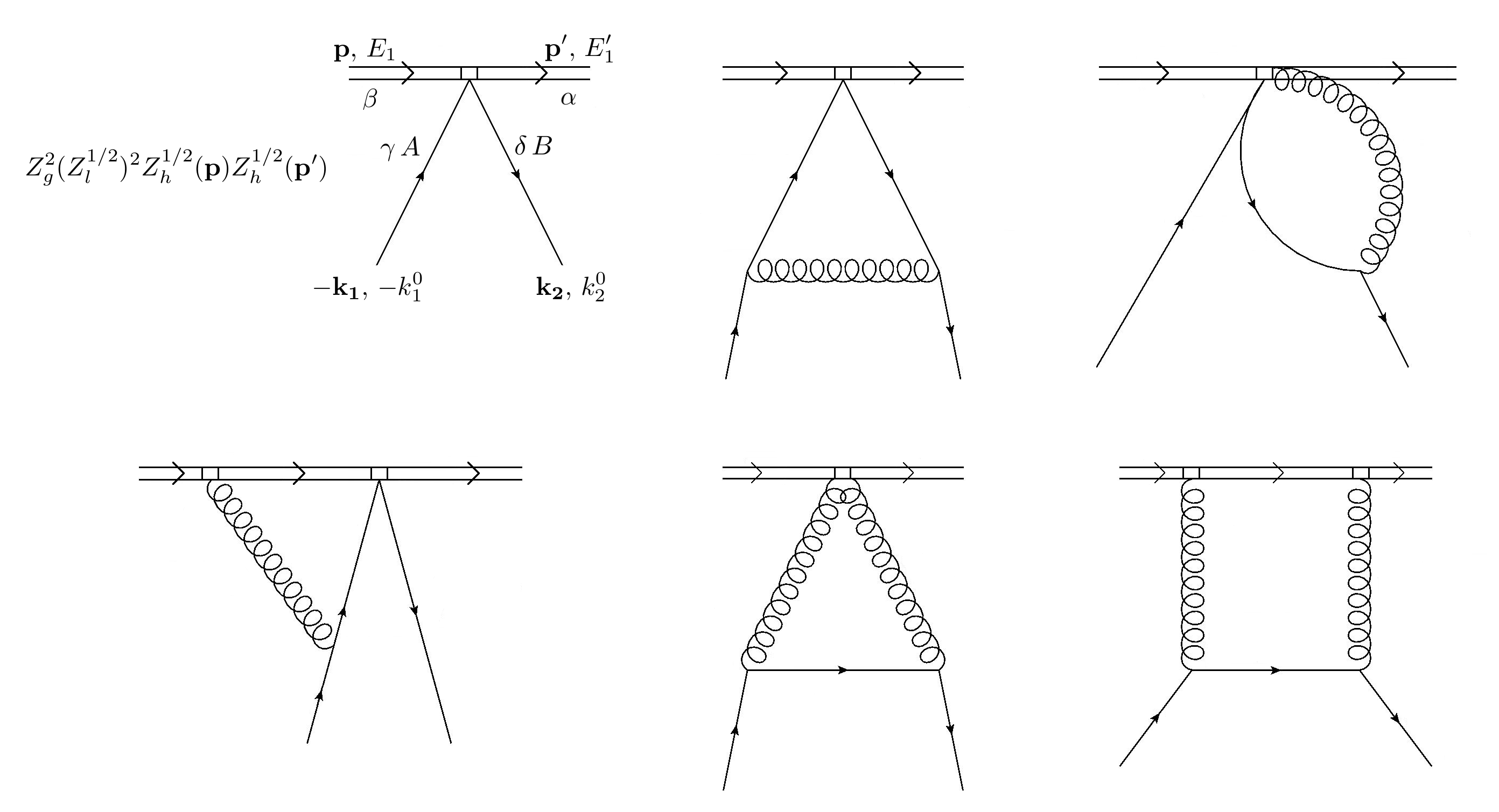}
	%
	%
	%
	%
\caption{ In the first diagram the counterterm of the external propagators and $g^2$ are included. The other diagrams are one loop topologies contributing to the anomalous dimensions of the $1/m^3$ Wilson coefficients in QCD of the heavy-light operators. All possible vertices and insertions with the right counting in $1/m$ should be considered to generate the diagrams. In general, the gluon propagator may be longitudinal or transverse.
\label{PlotsHeavyLight}}   
\end{figure}

Before going to the complete QCD case, we need to study with some detail the inclusion of light-fermions to the computation. The coefficients $c^{ll}$ and $c_D^l$ are NLL and we will neglect them in the following. 
The $c_i^{hl}$ were computed with LL accuracy in \cite{Bauer:1997gs}. The $d_i^{hl}$ are zero at LO at the hard scale (as there were the $c_i^{hl}$). There are no tree level contributions to these operators. The only way they can get nonzero LL running is through mixing with non-vanishing operators. The way we determine such mixing is by computing the divergent contributions to the elastic scattering of a heavy with a light quark. The divergences get then absorbed in the $d_i^{hl}$. As we already anticipated in previous sections not all 
Wilson coefficients get nonzero anomalous dimensions, but only a combination of some of them. We see in the calculation of the LL running 
of these operators that only three 
different structures appear. All the other possible structures are irrelevant for our calculation because its associated Wilson coefficient does not run and the matching condition i.e. the Wilson coefficient evaluated at the hard scale is zero, at least they are $\mathcal{O}(\alpha)$. These two properties together make the contribution of these operators subleading. The three operators needed are:
\begin{equation}
\label{O1}
 \mathcal{O}_1=\mathcal{M}_{4+}^{(3h)o} + \mathcal{M}_{4-}^{(3h)o}
 \,,
\end{equation}
\begin{equation}
 \mathcal{O}_2=\mathcal{M}_{3+}^{(3l)o} + \mathcal{M}_{3-}^{(3l)o}
 \,,
\end{equation}
\begin{equation}
 \mathcal{O}_3= \mathcal{M}_{3+}^{(3l)s} + \mathcal{M}_{3-}^{(3l)s}
 \,.
\end{equation}
The running of these operators is determined from the diagrams (topologies) drawn in Fig.~\ref{PlotsHeavyLight}. They produce around 57 diagrams to be computed (without counting 
crossed ones).

We will see later that $d_1^{hl}$ does not contribute to the running of the chromo-polarizabilities. Therefore, we will not consider it. For the other two Wilson coefficients we obtain
\bea
d_{2,B}^{hl}&=&d_2^{hl}-
\frac{1}{\epsilon}\frac{\al}{\pi}\left(\frac{2}{3}C_F-\frac{17}{24}C_A+\frac{\beta_0}{4}\right)d_2^{hl} +0d_{1,3}^{hl}
+
\frac{1}{\epsilon}\frac{\al}{\pi}
\left(8C_F-3C_A\right)
\left[
\right.
\nn
\\
&&
+
\left[0c_{A_1}-\frac{1}{192}\bar c_{A_2}+0c_{A_{3,4}}+\frac{1}{32}c_4+
0c_M
\right]
\nn
\\
&&
+
\left[\frac{1}{96}c_Sc_F+0 c_Dc_k
\right]
\nn
\\
&&
\left.
+
\left[
-\frac{5}{96}c_k^3
-\frac{5}{48}c_k c_F^2+\frac{1}{32}\bar c_{1}^{hl}c_k
\right]
\right]
\nn
\\
&&
+
\frac{1}{\epsilon}\frac{\al}{\pi}
\left[
+\frac{1}{8}c_{3}^{hl}c_k+C_A\frac{1}{32}c_{2}^{hl}c_F+0c_{4}^{hl}c_F
\right]
\,,\eea

\bea
d_{3,B}^{hl}&=&d_3^{hl} -\frac{1}{\epsilon}\frac{\al}{\pi}
\left(\frac{2}{3}C_F+\frac{\beta_0}{4}\right)d_3^{hl}+0d_{1,2}^{hl}
+\frac{1}{\epsilon}\frac{\al}{\pi}C_F(C_A-2C_F)\left[
\right.
\nn
\\
&&
+
\left[0c_{A_1}-\frac{1}{96}\bar c_{A_2}+0c_{A_3}-\frac{1}{96}c_{A_4}
+\frac{1}{16}c_4+0c_M
\right]
\nn
\\
&&
+
\left[\frac{1}{48}c_Sc_F+0 c_Dc_k
\right]
\nn
\\
&&
+
\left[
-\frac{5}{48}c_k^3-\frac{5}{24}c_k c_F^2
\right]
\nn
\\
&&
\left.
+\frac{1}{16}\bar c_{1}^{hl}c_k+0c_{3}^{hl}c_k+0c_{2,4}^{hl}c_F
\right]
\,,
\eea
which produce the following RG equations:
\bea
\nu\frac{d}{d\nu}d_2^{hl}&=&
-2\frac{\al}{\pi}\Bigg[
-
\left(\frac{2}{3}C_F-\frac{17}{24}C_A+\frac{\beta_0}{4}\right)d_2^{hl} 
\nn
\\
&&
+
\left(8C_F-3C_A\right)
\left[
-\frac{1}{192}\bar c_{A_2}+\frac{1}{32}c_4
+\frac{1}{96}c_Sc_F
-\frac{5}{96}c_k^3
-\frac{5}{48}c_kc_F^2+\frac{1}{32}\bar c_{1}^{hl}c_k
\right]
\nn
\\
&&
+\frac{1}{8}c_{3}^{hl}c_k+C_A\frac{1}{32}c_{2}^{hl}c_F
\Bigg]
\,,
\eea

\bea
&&
\nu\frac{d}{d\nu}d_3^{hl}=
-2\frac{\al}{\pi}
\Bigg[
-
\left(\frac{2}{3}C_F+\frac{\beta_0}{4}\right)d_3^{hl}
\\
\nn
&&
+C_F(C_A-2C_F)\left[
-\frac{1}{96}\bar c_{A_2}-\frac{1}{96}c_{A_4}
+\frac{1}{16}c_4+
\frac{1}{48}c_Sc_F
-\frac{5}{48}c_k^3-\frac{5}{24}c_k c_F^2
+\frac{1}{16}\bar c_{1}^{hl}c_k
\right]
\Bigg]
\,.
\eea

Quite remarkably, the RG equations depend only on gauge-independent combinations of Wilson coefficients: $\bar c_{A_2}$ and $\bar c_1^{hl}$. 
This is quite a strong check, as at intermediate steps we get contributions from $c_D$, $c_M$, $c_{A_2}$, $c_1^{hl}$, and $d_1^{hl}$, which only at the end of the computation combine in gauge-independent sets.

The other $d$ coefficients have the structure ($i,j > 3$)
\be
\nu\frac{d}{d\nu}d_i^{hl}=\frac{\al}{\pi}A_{i,j}d_j^{hl}
\,.
\ee
Therefore, they are NLL.

\subsection{QCD case}
\label{sec:RGeq}

For the pure gluonic sector we have that $c_1^g$ is NLL. Therefore, we will neglect it in the following. 

\begin{figure}[!htb]
	\includegraphics[width=0.95\textwidth]{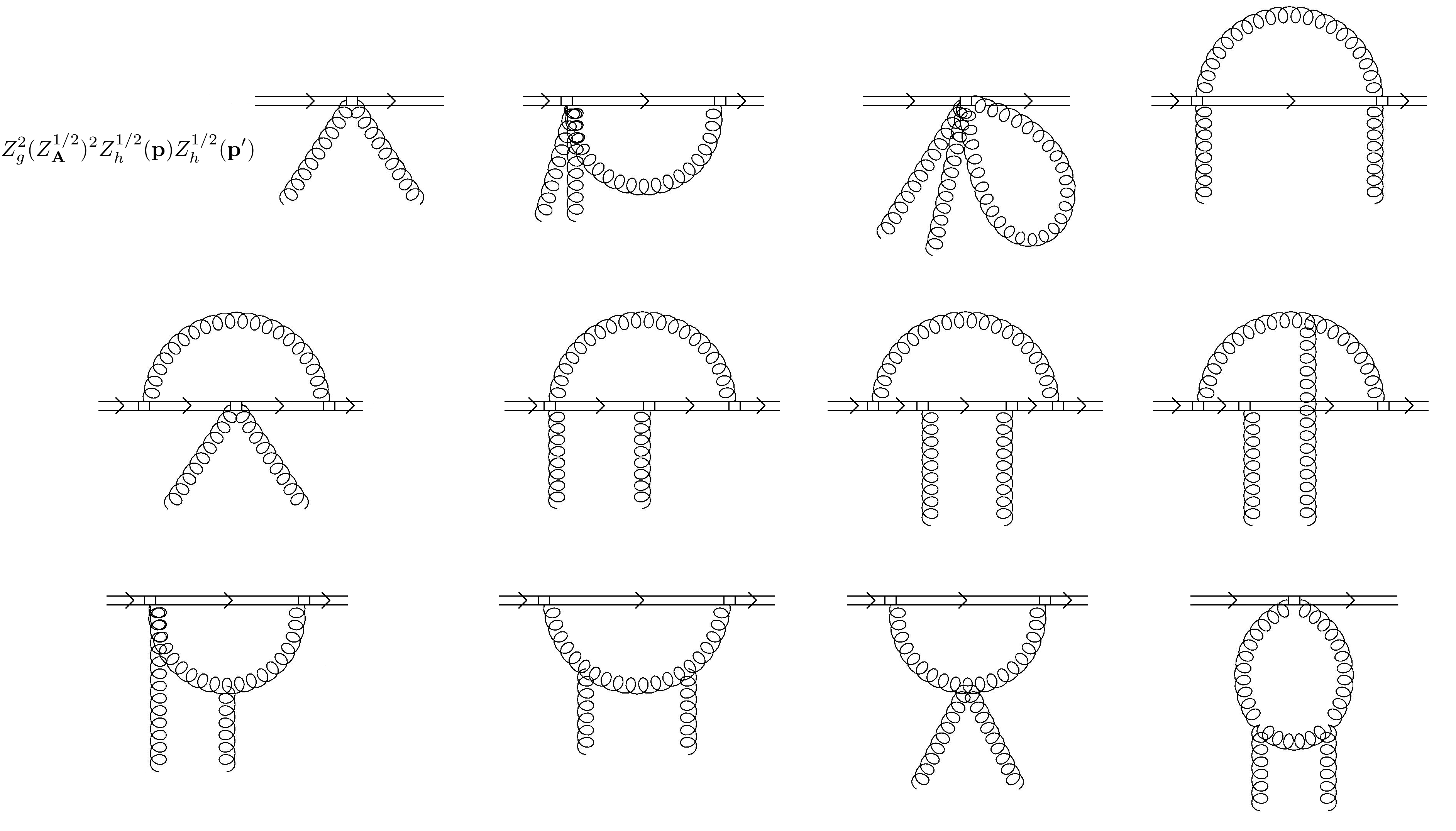}
	\includegraphics[width=0.95\textwidth]{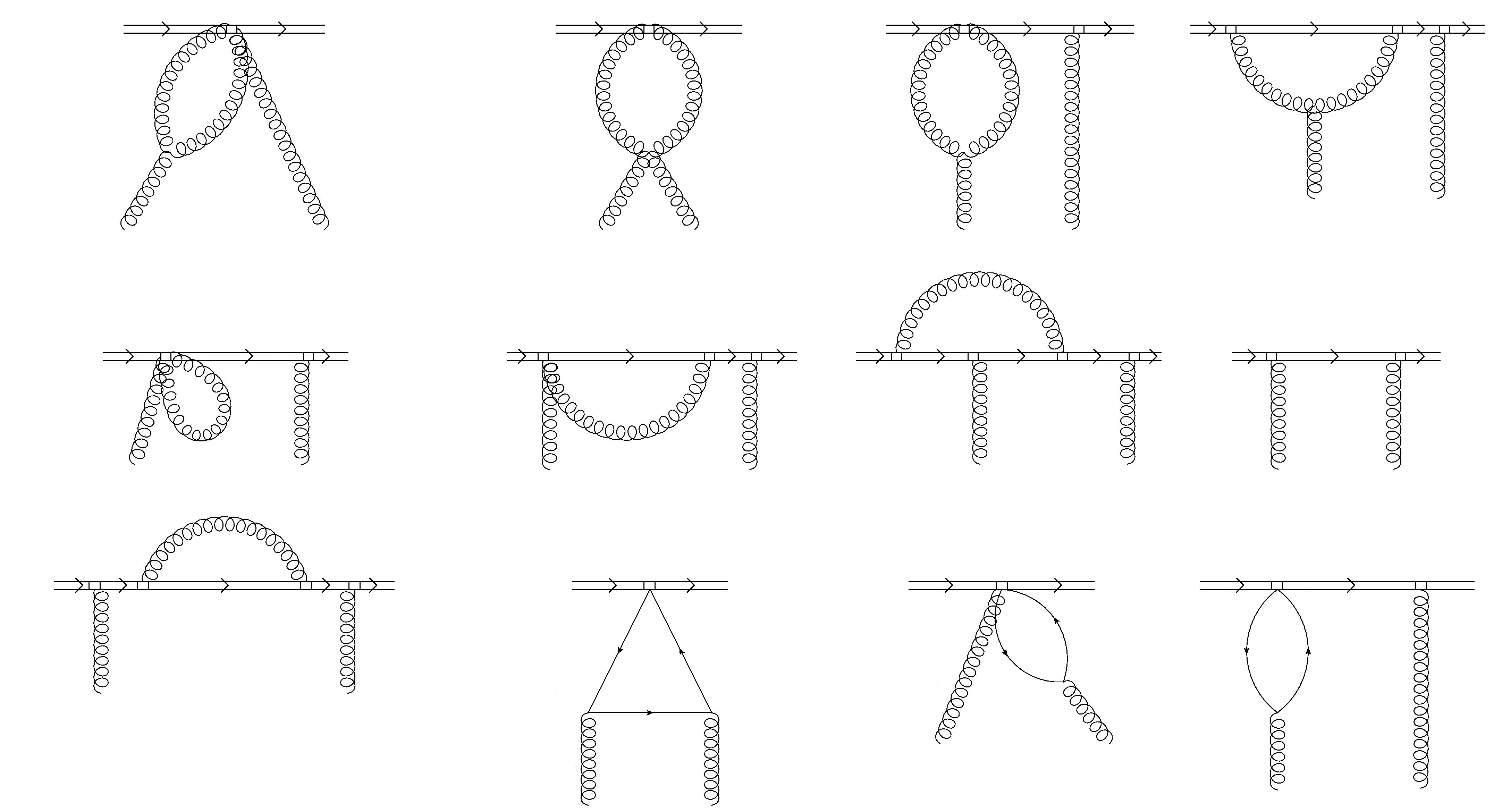}
	%
	%
%
	%
	%
	%
\caption{The tree level diagrams drawn should be understood to be multiplied by counterterms: either from low order in $1/m$, from external wave function counterterms, or from $g$. The remaining diagrams are the one loop topologies contributing to the anomalous dimensions of the $1/m^3$ Wilson coefficients in QCD of the heavy-gluon operators. Final diagrams are generated from those topologies by considering all possible vertices and kinetic insertions to ${\cal O}(1/m^3)$.
\label{PlotsHeavygluon1}}   
\end{figure}

The running of $\{c_{A_i},c_4\}$ is determined from the topologies drawn in Fig.~\ref{PlotsHeavygluon1}. Out of these topologies we generate all possible diagrams of order $1/m^3$ by considering all possible vertices to the appropriate order in $1/m$ and/or kinetic insertions. This generates around 200 diagrams (without taking into account permutations and crossing). These topologies refer to the elastic scattering of the heavy quark with a transverse gluon. The topologies of the elastic scattering of the heavy quark with a longitudinal gluon
(which we also compute) are the same (though, obviously, the external wave function counterterms change) and also generate around 200 diagrams (without taking into account permutations and crossing). Note that each of the internal gluon propagators 
may refer to a transverse or longitudinal gluon. 

	%
	%
%
	%
	%
	%

Since we work in the massless limit the expression for $d_i^{hl}$ does not depend on the specific light fermion, so the result of each diagram has to be multiplied by the number of light fermions. 

For diagrams proportional to $1/m^3$ operators only irreducible diagrams need to be considered. When one considers diagrams proportional to iterations of $1/m^2$ and/or $1/m$ operators one also has to consider reducible diagrams, following the same line of argumentation as for the discussion of the QED case.

Overall, we obtain
\bea
c_{4,B}&=&c_{4}+0 c_{A_i}+0 c_4+0 c_M
+0 c_Dc_k +0c_Sc_F +0c_k^3+0c_F^3+0c_kc_F^2+0c_{1,3}^{hl}c_k+0c_{2,4}^{hl}c_F+0d_{1,2,3}^{hl}
\,,
\nn
\\
\\
c_{A_1,B}&=&c_{A_1}-\frac{2}{3}c_{A_1}
C_A \frac{\al}{\pi}\frac{1}{\epsilon}+\frac{1}{2}c_4C_A \frac{\al}{\pi}\frac{1}{\epsilon}
+\left[\frac{5}{12}c_{A_1}
+
\frac{11}{96}\bar c_{A_2}
\right]C_A
\frac{\al}{\pi}\frac{1}{\epsilon}+0c_{A_{3,4}}
+0c_M
\nn
\\
&&+0 c_Dc_k+\frac{1}{48}c_Sc_F
C_A \frac{\al}{\pi}\frac{1}{\epsilon}
\nn
\\
&&-\frac{1}{6}c_k^3(16C_F+11C_A) \frac{\al}{\pi}\frac{1}{\epsilon}
-\frac{3}{4}c_F^3C_A \frac{\al}{\pi}\frac{1}{\epsilon}
-\frac{73}{48}c_F^2c_kC_A \frac{\al}{\pi}\frac{1}{\epsilon}
\nn
\\
&&
+0c_{1,3}^{hl}c_k+0c_{2,4}^{hl}c_F+0d_1^{hl}+\frac{8}{3}T_Fn_f  \frac{\al}{\pi}\frac{1}{\epsilon}d_2^{hl}+0d_3^{hl}
\,,
\\
\bar c_{A_2,B}&\equiv&c_{A_2,B}-4c_{M,B}=
\bar c_{A_2}-\frac{11}{24}\bar c_{A_2}
C_A \frac{\al}{\pi}\frac{1}{\epsilon}
-4c_4
C_A \frac{\al}{\pi}\frac{1}{\epsilon}
+
\frac{1}{2}c_{A_1}
C_A \frac{\al}{\pi}\frac{1}{\epsilon}
+0c_{A_{3,4}}+0c_M
\nn
\\
&&
+0c_Dc_k
C_A \frac{\al}{\pi}\frac{1}{\epsilon}-\frac{1}{12}c_Sc_F
C_A \frac{\al}{\pi}\frac{1}{\epsilon}
\nn
\\
&&+\frac{4}{3}c_k^3(8C_F+7C_A) \frac{\al}{\pi}\frac{1}{\epsilon}
+\frac{3}{2}c_F^3C_A \frac{\al}{\pi}\frac{1}{\epsilon}
+\frac{43}{12}c_F^2c_kC_A \frac{\al}{\pi}\frac{1}{\epsilon}
\nn
\\
&&+0c_{1,3}^{hl}c_k+0c_{2,4}^{hl}c_F+0d_1^{hl}
-\frac{32}{3}T_Fn_f  \frac{\al}{\pi}\frac{1}{\epsilon}d_2^{hl}+0d_3^{hl}
\,,
\\
c_{A_3,B}&=&c_{A_3}-\frac{2}{3}c_{A_3}C_A \frac{\al}{\pi}\frac{1}{\epsilon}
+
\left[
\frac{1}{4}c_{A_1}
+
\frac{11}{96}\bar c_{A_2}
+
\frac{2}{3}c_{A_3}
+
\frac{11}{48}c_{A_4}
\right]
C_A\frac{\al}{\pi}\frac{1}{\epsilon}
+0c_M
\nn
\\
&&+0 c_Dc_k+\frac{1}{48}c_Sc_F
C_A \frac{\al}{\pi}\frac{1}{\epsilon}
\nn
\\
&&
+0c_k^3+\frac{3}{4}c_F^3C_A \frac{\al}{\pi}\frac{1}{\epsilon}
-\frac{73}{48}c_F^2c_kC_A \frac{\al}{\pi}\frac{1}{\epsilon}
\nn
\\
&&
+0c_{1,3}^{hl}c_k+0c_{2,4}^{hl}c_F+0d_1^{hl}-\frac{8}{3}T_Fn_f \frac{\al}{\pi}\frac{1}{\epsilon}d_2^{hl}+\frac{16}{3}T_Fn_f \frac{\al}{\pi}\frac{1}{\epsilon}C_A d_3^{hl}
\,,
\\
c_{A_4,B}&=&c_{A_4}-\frac{2}{3}c_{A_4}
C_A \frac{\al}{\pi}\frac{1}{\epsilon}
-\frac{4}{3}c_4C_A \frac{\al}{\pi}\frac{1}{\epsilon}
-
\left[
\frac{1}{2}c_{A_1}
+
\frac{11}{24}\bar c_{A_2}+0c_{A_3}
+
\frac{1}{4}c_{A_4}
\right]
C_A
\frac{\al}{\pi}\frac{1}{\epsilon}
+
0c_M
\nn
\\
&&
+0c_Dc_k
C_A \frac{\al}{\pi}\frac{1}{\epsilon}
+\frac{17}{12}c_Sc_F
C_A \frac{\al}{\pi}\frac{1}{\epsilon}
\nn
\\
&&+\frac{4}{3}c_k^3C_A \frac{\al}{\pi}\frac{1}{\epsilon}
-\frac{3}{2}c_F^3C_A \frac{\al}{\pi}\frac{1}{\epsilon}
+\frac{31}{12}c_F^2c_kC_A \frac{\al}{\pi}\frac{1}{\epsilon}
\nn
\\
&&+0c_{1,3}^{hl}c_k+0c_{2,4}^{hl}c_F+0d_1^{hl}+\frac{32}{3}T_Fn_f  \frac{\al}{\pi}\frac{1}{\epsilon}d_2^{hl}-\frac{64}{3}T_Fn_f  \frac{\al}{\pi}\frac{1}{\epsilon}C_A d_3^{hl}
\,.
\eea
This produces the following RG equations: 
\bea
\nu\frac{d}{d\nu}c_{4}&=&0
\,,
\\
\nn
\nu\frac{d}{d\nu}c_{A_1}&=&
-2\frac{\al}{\pi}\Bigg[
 +\frac{1}{2}c_4C_A 
+\left[-\frac{1}{4}c_{A_1}
+
\frac{11}{96}\bar c_{A_2}
\right]C_A
\nn
\\
&&-\frac{1}{6}c_k^3(16C_F+11C_A) 
-\frac{3}{4}c_F^3C_A 
-\frac{73}{48}c_F^2c_kC_A 
+\frac{1}{48}c_Sc_F
C_A 
+\frac{8}{3}T_Fn_f  d_2^{hl}
\Bigg]
\,,
\\
\nu\frac{d}{d\nu}\bar c_{A_2}&=&
-2 \frac{\al}{\pi}\Bigg[
-\frac{11}{24}\bar c_{A_2}
C_A 
-4c_4
C_A 
+
\frac{1}{2}c_{A_1}
C_A
\nn
\\
&&
-\frac{1}{12}c_Sc_F
C_A
+\frac{4}{3}c_k^3(8C_F+7C_A) 
+\frac{3}{2}c_F^3C_A
+\frac{43}{12}c_F^2c_kC_A 
-\frac{32}{3}T_Fn_f  d_2^{hl}
\Bigg]
\,,
\\
\nn
\nu\frac{d}{d\nu}c_{A_3}&=&
-2\frac{\al}{\pi}
\Bigg[
+
\left[
\frac{1}{4}c_{A_1}
+
\frac{11}{96}\bar c_{A_2}
+
\frac{11}{48}c_{A_4}
\right]
C_A
\nn
\\
&&+\frac{1}{48}c_Sc_F
C_A 
+\frac{3}{4}c_F^3C_A
-\frac{73}{48}c_F^2c_kC_A 
-\frac{8}{3}T_Fn_f d_2^{hl}+\frac{16}{3}T_Fn_f C_A d_3^{hl}
\Bigg]
\,,
\\
\nn
\nu\frac{d}{d\nu}c_{A_4}&=&
-2\frac{\al}{\pi}
\Bigg[
-\frac{4}{3}c_4C_A 
-
\left[
\frac{1}{2}c_{A_1}
+
\frac{11}{24}\bar c_{A_2}
+
\frac{11}{12}c_{A_4}
\right]
C_A
\nn
\\
&&
+\frac{17}{12}c_Sc_F
C_A 
+\frac{4}{3}c_k^3C_A 
-\frac{3}{2}c_F^3C_A 
+\frac{31}{12}c_F^2c_kC_A 
+\frac{32}{3}T_Fn_f  d_2^{hl}-\frac{64}{3}T_Fn_f  C_A d_3^{hl}
\Bigg]
\,.
\nn
\\
\eea
It is worth mentioning that the running of $c_{A_1}$ and $\bar c_{A_2}$ can be obtained without determining the running of $c_{A_3}$ and $c_{A_4}$.

Again, it is quite remarkable that the RG equations depend only on gauge-independent combinations of Wilson coefficients: $\bar c_{A_2}$ and $\bar c_1^{hl}$. 
This is quite a strong check, as at intermediate steps we get contributions from $c_D$, $c_M$, $c_{A_2}$, $c_1^{hl}$, and $d_1^{hl}$, which only at the end of the computation combine in gauge-independent sets.

\section{Solution and numerical analysis}

The RG equations obtained in the previous section depend on a list of 7 Wilson coefficients ${\bf A}=\{ c_{A_1}, \bar c_{A_2}, c_{A_3}, c_{A_4},c_4,d_2^{hl},d_3^{hl} \}$. The last two are only computed since they are nonzero at LL and contribute to the different $c_{A_i}$. The running of $c_4$ is zero, as predicted by reparameterization invariance. 
The remaining equations can be written, more compactly, in a matricial form 
\be
\nu \frac{d}{d \nu} {\bf A}= \frac{\al}{\pi}{\bf M}  {\bf A}+{\bf f}(\al)
\,.
\ee
The matrix ${\bf M}$ follows from the results of the previous section. We only need the LL running of $\al$:
\be
 \nu   {{\rm d} \over {\rm d} \nu}   \al
\equiv
\beta(\alpha_s)
=
-2 \al
\left\{\beta_0{\al \over 4 \pi} + \cdots\right\}
\,,
\ee
where 
\begin{eqnarray}
  \beta_0&=&{11 \over 3}C_A -{4 \over 3}T_Fn_f\,,
\end{eqnarray}
and $n_f$ is the number of dynamical (active) quarks. 
 
Then the above equation can be simplified to
\be
\frac{d}{d \al} {\bf A}= -\frac{2}{ \beta_0\al}{\bf M}  {\bf A}- \frac{2\pi}{ \beta_0\al^2}{\bf f}(\al)
\,.
\ee
 
We also need the initial matching conditions at the hard scale. The tree-level Wilson coefficients at the hard scale have been determined in Ref.~\cite{Manohar:1997qy}. They read $c_k=c_4=c_F=c_D=c_S=c_{A_1}=1$
and
$c_M=c_{A_2}=c_{A_3}=c_{A_4}=0$.

Please note that the matching coefficients of the
kinetic term are protected by reparameterization invariance
($c_k=c_4=1$ to any order in perturbation theory) \cite{Luke:1992cs}.
Nevertheless, even if we set them to 1 when solving the RG equations, we have kept them explicit in the RG equations for
tracking purposes. 

The initial matching conditions of the $d_i^{hl}$ and $c_i^{hl}$ are ${\cal O}(\al)$ unless the operators can be generated at tree level. This is not the case with the basis we consider, but it is if one eliminates the Darwin operator $c_D$ (i.e. for $\bar c_1^{hl}$).

After solving the RG  equations we obtain the running of the different Wilson coefficients. For the case without light fermions, we can obtain analytic results. They read ($z=\left[\frac{\al(\nu)}{\al(m)}\right]^{\frac{1}{\beta_0}}\simeq
1-\frac{1}{2\pi}\al(\nu)\ln (\frac{\nu}{m})$) 
\bea
c_{A_1}&=&+\frac{75}{17} z^{-3 C_A}-\frac{29}{3} z^{-2
   C_A}-\frac{z^{-C_A}}{11}
   \nn
   \\
   &&+
   \left(
   \frac{64 C_F }{\sqrt{157}
   C_A}+\frac{42184 }{561
   \sqrt{157}}+\frac{1780}{561} 
   \right)z^{-\frac{1}{12}
   \left(17+\sqrt{157}\right) C_A}
   \nn
   \\
   &&
  +\left( -\frac{64 C_F }{\sqrt{157} C_A}-\frac{42184 }{561
   \sqrt{157}}+\frac{1780}{561}
   \right) z^{\frac{1}{12}
   \left(\sqrt{157}-17\right) C_A}
   \,,
\eea
\bea
\bar c_{A_2}&=&
-\frac{216}{17} z^{-3 C_A}+34 z^{-2 C_A}+\frac{2
   z^{-C_A}}{11}+\frac{128}{11}+\frac{256 C_F}{11
   C_A}
   \nn
   \\
   &&
   -\left(\frac{640 C_F}{11
   \sqrt{157} C_A}+\frac{128 C_F }{11 C_A}+\frac{29720 }{187
   \sqrt{157}}+\frac{3096}{187} 
   \right)z^{-\frac{1}{12}
   \left(17+\sqrt{157}\right) C_A}
      \nn
   \\
   &&
+\left(\frac{640 C_F }{11 \sqrt{157} C_A}-\frac{128 C_F }{11 C_A}
   +\frac{29720 }{187
   \sqrt{157}}-\frac{3096}{187}
   \right) z^{\frac{1}{12}
   \left(\sqrt{157}-17\right) C_A}
   \,,
\eea
\bea
c_{A_3}&=&
\left(\frac{32 C_F }{11 C_A}+\frac{344}{55}\right) z^{-\frac{11
   C_A}{3}}-\frac{75}{17} z^{-3 C_A}+\frac{88}{15} z^{-2
   C_A}+\frac{23 z^{-C_A}}{11}-\frac{38}{11}-\frac{32
   C_F}{11 C_A}
   \nn
   \\
   &&
 - \left( \frac{64 C_F }{\sqrt{157} C_A}
   +\frac{42184 }{561
   \sqrt{157}}+\frac{1780}{561} \right)z^{-\frac{1}{12}
   \left(17+\sqrt{157}\right) C_A}
 \nn
   \\
   &&+
   \left(\frac{64 C_F
   }{\sqrt{157}
   C_A}+\frac{42184 }{561
   \sqrt{157}}-\frac{1780}{561} 
   \right)z^{\frac{1}{12}
   \left(\sqrt{157}-17\right) C_A}
   \,,
     \eea
\bea
c_{A_4}&=&
   -\left(\frac{128 C_F}{11 C_A}+\frac{1376}{55}\right) z^{-\frac{11 C_A}{3}}+\frac{216}{17}
   z^{-3 C_A}-\frac{64}{5} z^{-2 C_A}-\frac{24
   z^{-C_A}}{11}-\frac{64}{11}-\frac{128 C_F}{11
   C_A}
   \nn
   \\
   &&
  +
  \left(\frac{640 C_F
  }{11
   \sqrt{157} C_A} +\frac{128 C_F}{11 C_A}+\frac{29720 }{187
   \sqrt{157}}+\frac{3096}{187} 
   \right)z^{-\frac{1}{12}
   \left(17+\sqrt{157}\right) C_A}
 \nn
   \\
   &&
   +\left(
   -\frac{640 C_F
   }{11
   \sqrt{157} C_A}
   +\frac{128 C_F }{11 C_A}-\frac{29720 }{187
   \sqrt{157}}
   +\frac{3096}{187}
   \right) z^{\frac{1}{12}
   \left(\sqrt{157}-17\right) C_A}
     \,.
   \eea
After the inclusion of light fermions, the solution of the RG equations is numerical. We show the result for $n_f=4$ light fermions where 
$\al(m)$ has 
$n_f$ active light flavors:
\bea
c_{A_1}&=&
   1.08839\times10^{-9} + \frac{5.9421}{z^{9.}} 
- \frac{3.66729\times 10^{-19}}{z^{8.33333}} + \frac{3.44328}{z^{7.99055}} - \frac{2.07923}{z^{6.83333}} 
+ \frac{2.56988}{z^{6.5}}    
   \nn
   \\
   &&
    - \frac{2.31629}{z^{6.}}  + \frac{8.05227}{z^{3.}} - \frac{10.0312}{z^{2.87467}} 
- \frac{4.58085}{z^{1.02367}}     
\,,
\eea
\bea
\bar c_{A_2}&=&
   23.9049 - \frac{3.83509\times 10^{-20}}{z^{14.5556}} - \frac{17.807}{z^{9.}} 
+ \frac{4.00069\times10^{-19}}{z^{8.33333}} - \frac{11.1876}{z^{7.99055}}  + \frac{6.49126}{z^{6.83333}}    
   \nn
   \\
   &&
     - \frac{7.90732}{z^{6.5}} + \frac{11.9489}{z^{6.}} - \frac{16.1045}{z^{3.}} + \frac{19.1876}{z^{2.87467}}
- \frac{8.5262}{z^{1.02367}}    
\,,
\eea
\bea
c_{A_3}&=&
    -4.73174 + \frac{1.11951}{z^{14.5556}} - \frac{5.9421}{z^{9.}}  + \frac{12.652}{z^{8.33333}} 
- \frac{3.44328}{z^{7.99055}} + \frac{1.45822}{z^{6.83333}}  + \frac{1.5325608\times10^{-21}}{z^{41/6}}    
   \nn
   \\
   &&
    + \frac{0.621007}{z^{6.83333}} - \frac{5.10498}{z^{6.5}} + \frac{5.522246\times 10^{-21}}{z^{13/2}}  + \frac{1.5000000}{z^{6}} 
- \frac{6.68835}{z^{6.}}  + \frac{0.\times 10^{-28}}{z^{23/6}}     
   \nn
   \\
   &&
     + \frac{1.500000}{z^{3}} - \frac{7.55227}{z^{3.}}  
+ \frac{10.0312}{z^{2.87467}} + \frac{4.58085}{z^{1.02367}} +  0.\times 10^{-5} \ln {z^{1/6}}     
   \,,
   \eea
\bea
c_{A_4}&=&
     -12.9779 - \frac{4.47804}{z^{14.5556}} + \frac{17.807}{z^{9.}}  - \frac{ 50.6079}{z^{8.33333}} 
+ \frac{11.1876}{z^{7.99055}} 
 - \frac{6.49126}{z^{6.83333}}  + \frac{18.0477}{z^{6.5}}    
   \nn
   \\
   && 
    + \frac{24.0697}{z^{6.}}   
   + \frac{14.1045}{z^{3.}}
   - \frac{19.1876}{z^{2.87467}}  + \frac{8.5262}{z^{1.02367}}     
   \,,
   \eea
   \bea
d_2^{hl}&=&  
     -0.124078 - \frac{0.101624}{z^{9.}}  - \frac{0.084417}{z^{7.99055}}  + \frac{0.0354398}{z^{6.83333}}
   - \frac{0.0802099}{z^{6.83333}}  + \frac{0.0880307}{z^{6.5}}    
   \nn
   \\
   &&
    + \frac{0.387618}{z^{6.}} 
  + \frac{1.04971}{z^{3.}}  - \frac{1.29563}{z^{2.87467}}  + \frac{0.125167}{z^{1.02367}}     
   \,,
   \eea
   \bea
   d_3^{hl}&=&
    0.00135186 - \frac{0.0310975}{z^{14.5556}} 
+ \frac{1.63677\times10^{-19}}{z^{9.}} + \frac{0.263583}{z^{8.33333}} + \frac{1.0633\times10^{-19}}{z^{7.99055}}    
   \nn
   \\
   &&   
       + \frac{1.56003\times10^{-20}}{z^{6.83333}} - \frac{0.0891247}{z^{6.5}}  - \frac{0.144712}{z^{6.}} 
+ \frac{4.54348\times10^{-12}}{z^{3.}}     
   \nn
   \\
   &&   
   + \frac{1.68781\times10^{-19}}{z^{2.87467}}    
   - \frac{1.60402\times10^{-20}}{z^{1.02367}}      
\,.
\eea
Let us finally notice that the running of $c_{A_2}$ can be deduced from the above results using the expression of $c_M$ determined from reparameterization invariance, which in turn depends on $c_D$. The result will depend on the gauge though.

If we expand the above solutions in powers of $\al$, we can explicitly write the single log (it can also be obtained by trivial inspection of the RG equations in Sec. \ref{Sec:RGE}). We obtain
\bea
\label{cA1SL}
c_{A_1}&=&1+\left(\frac{16}{3}C_F+\frac{23}{3}C_A\right)\frac{\al}{\pi}\ln \left(\frac{\nu}{m}\right)
+{\cal O}(\al^2)
\,,
\\
\bar c_{A_2}&=&0-\left(\frac{64}{3}C_F+\frac{65}{3}C_A\right)\frac{\al}{\pi}\ln \left(\frac{\nu}{m}\right)
+{\cal O}(\al^2)
\,,
\\
\label{cA3SL}
c_{A_3}&=&0+C_A\frac{\al}{\pi}\ln \left(\frac{\nu}{m}\right) +{\cal O}(\al^2)
\,,
\\
\label{cA4SL}
c_{A_4}&=&0-4C_A\frac{\al}{\pi}\ln \left(\frac{\nu}{m}\right) +{\cal O}(\al^2)
\,,
\\
\label{d2hlSL}
d_2^{hl}&=&0+ \frac{1}{6}\left(8C_F-3C_A\right)
\frac{\al}{\pi}\ln \left(\frac{\nu}{m}\right)
+{\cal O}(\al^2)
\,,
\\
\label{d3hlSL}
d_3^{hl}&=&0+ \frac{1}{3}C_F\left(C_A-2C_F\right)
\frac{\al}{\pi}\ln \left(\frac{\nu}{m}\right)
+{\cal O}(\al^2)
\,.
\eea

\begin{figure}[!htb]
	\includegraphics[width=0.43\textwidth]{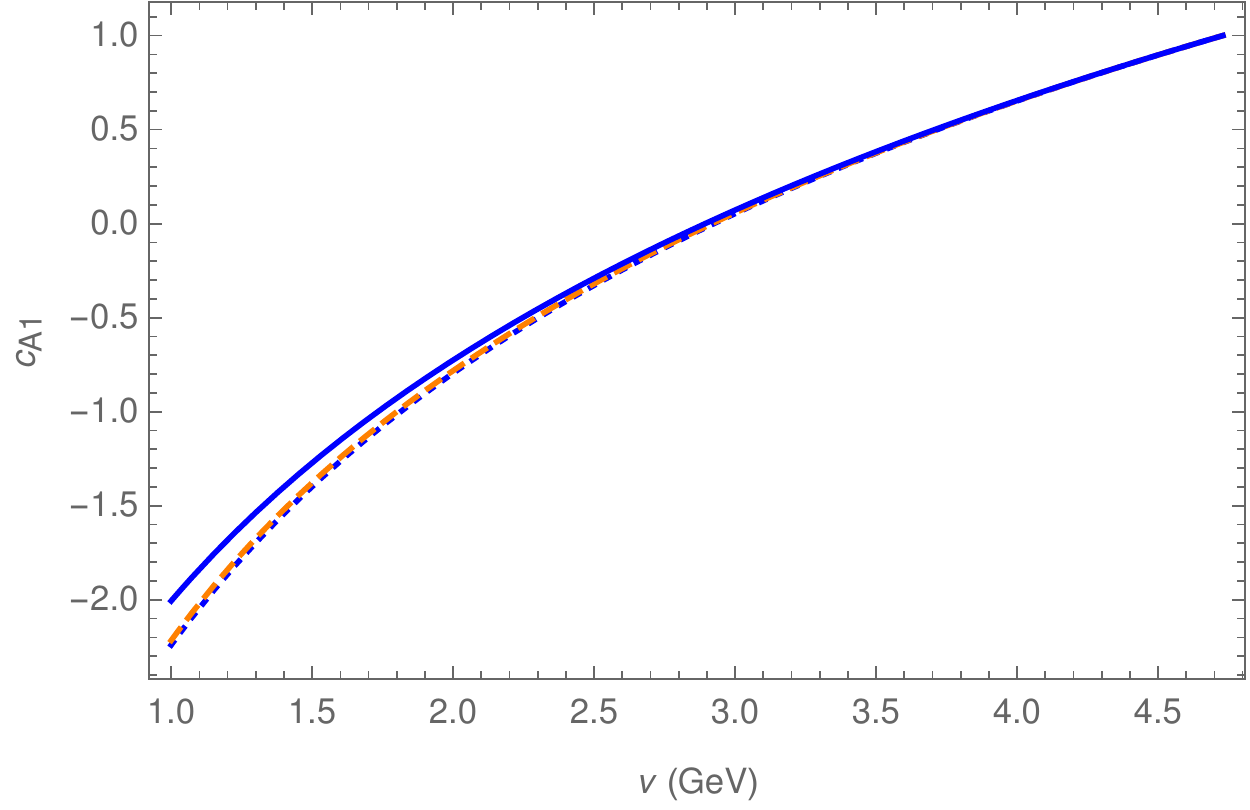}
\hspace{6ex}
	\includegraphics[width=0.43\textwidth]{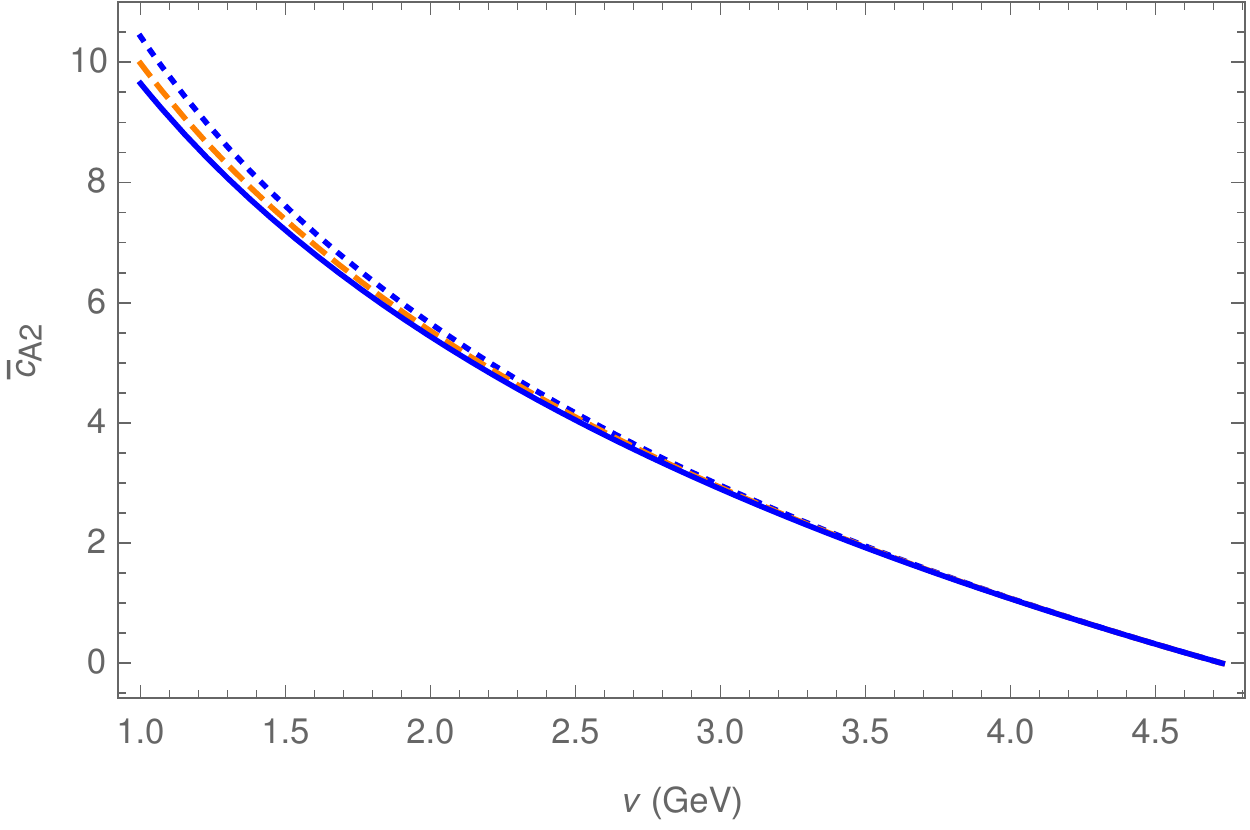}
	%
	\vspace{1ex}
	\includegraphics[width=0.43\textwidth]{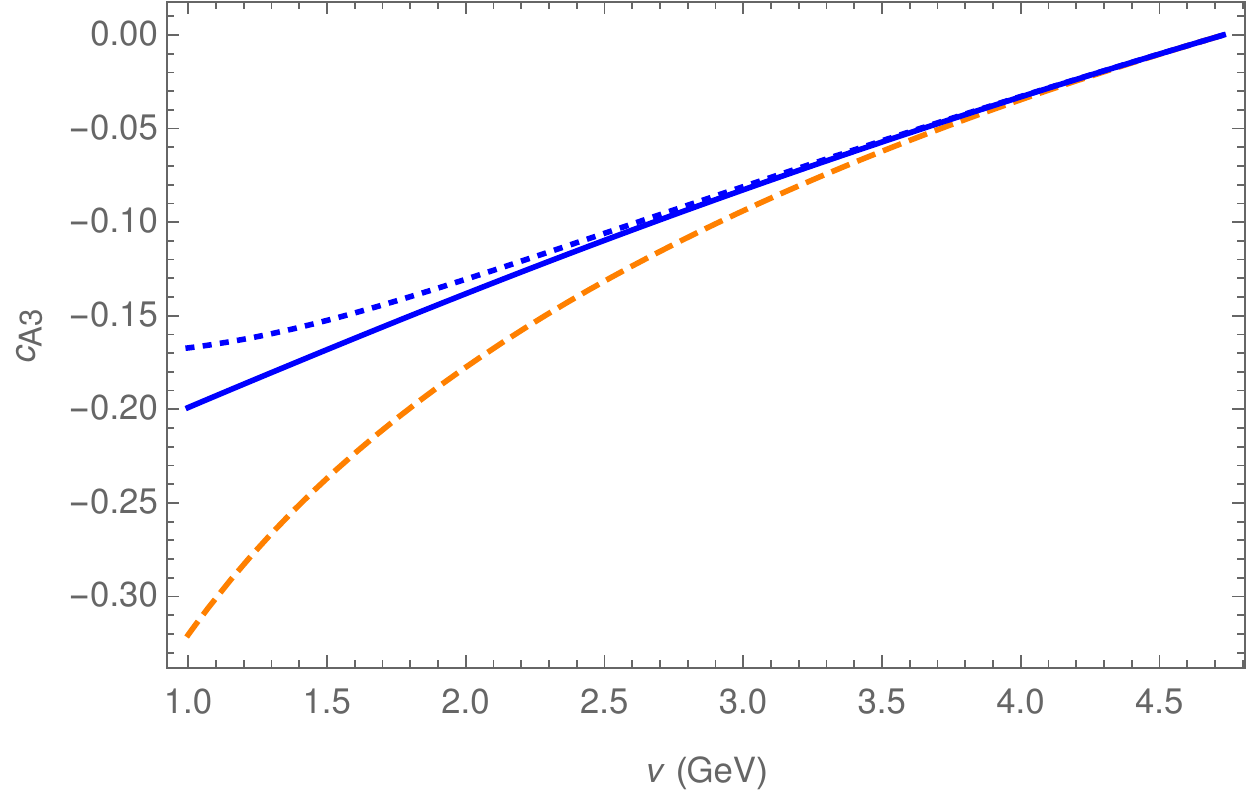}
\hspace{6ex}
	\includegraphics[width=0.43\textwidth]{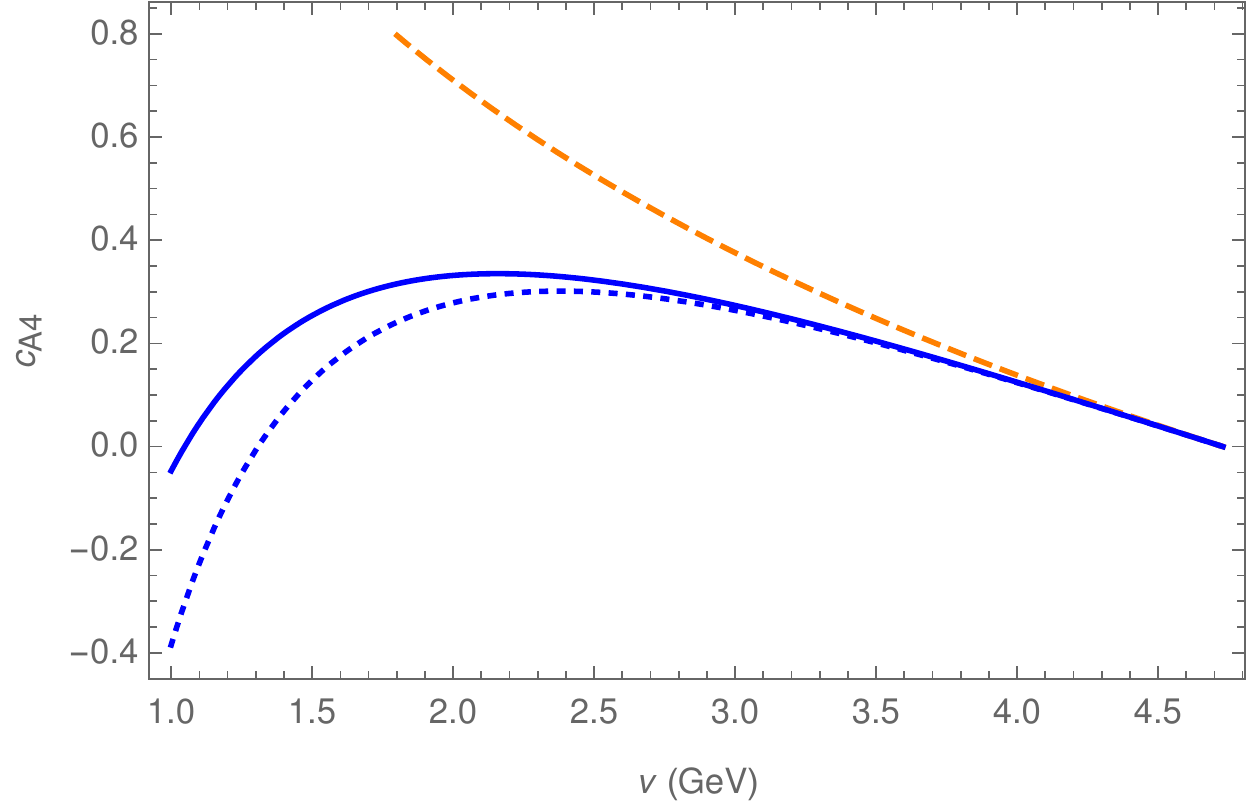}
	%
	\vspace{1ex}
	\includegraphics[width=0.43\textwidth]{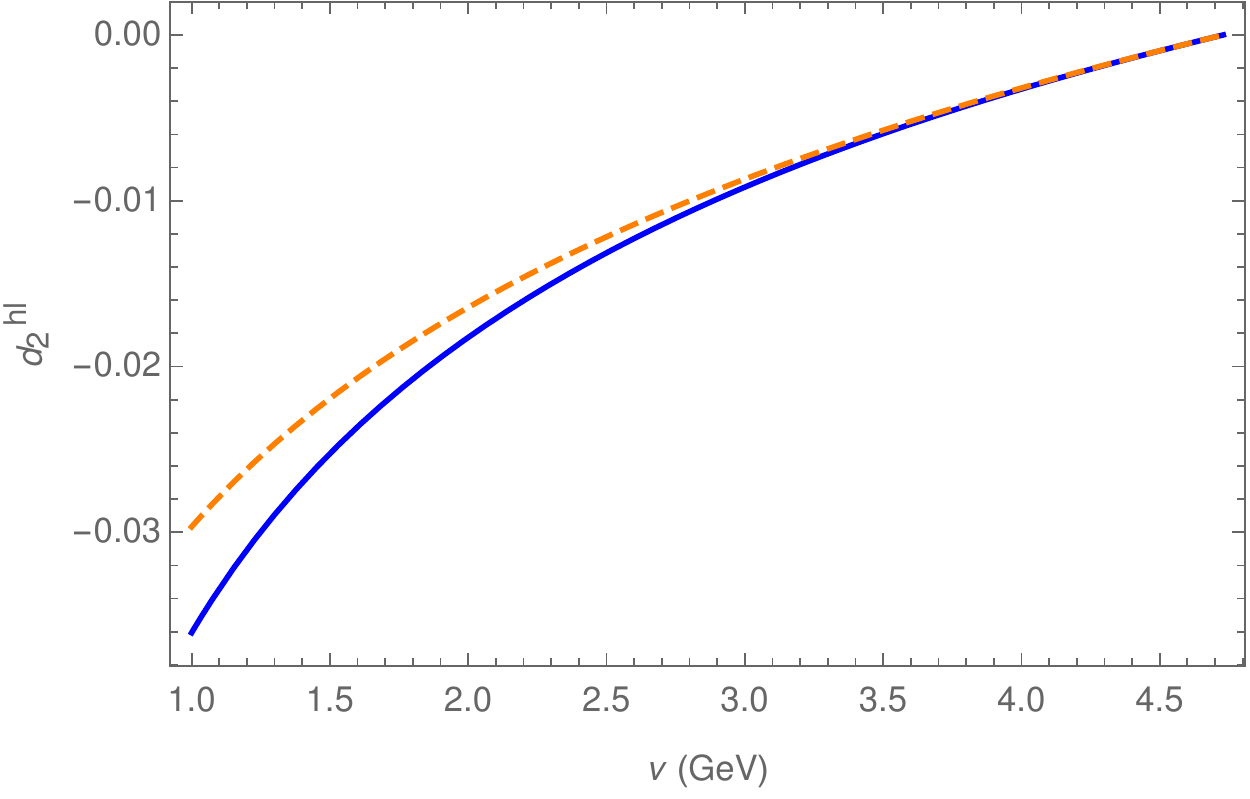}
\hspace{6ex}
	\includegraphics[width=0.43\textwidth]{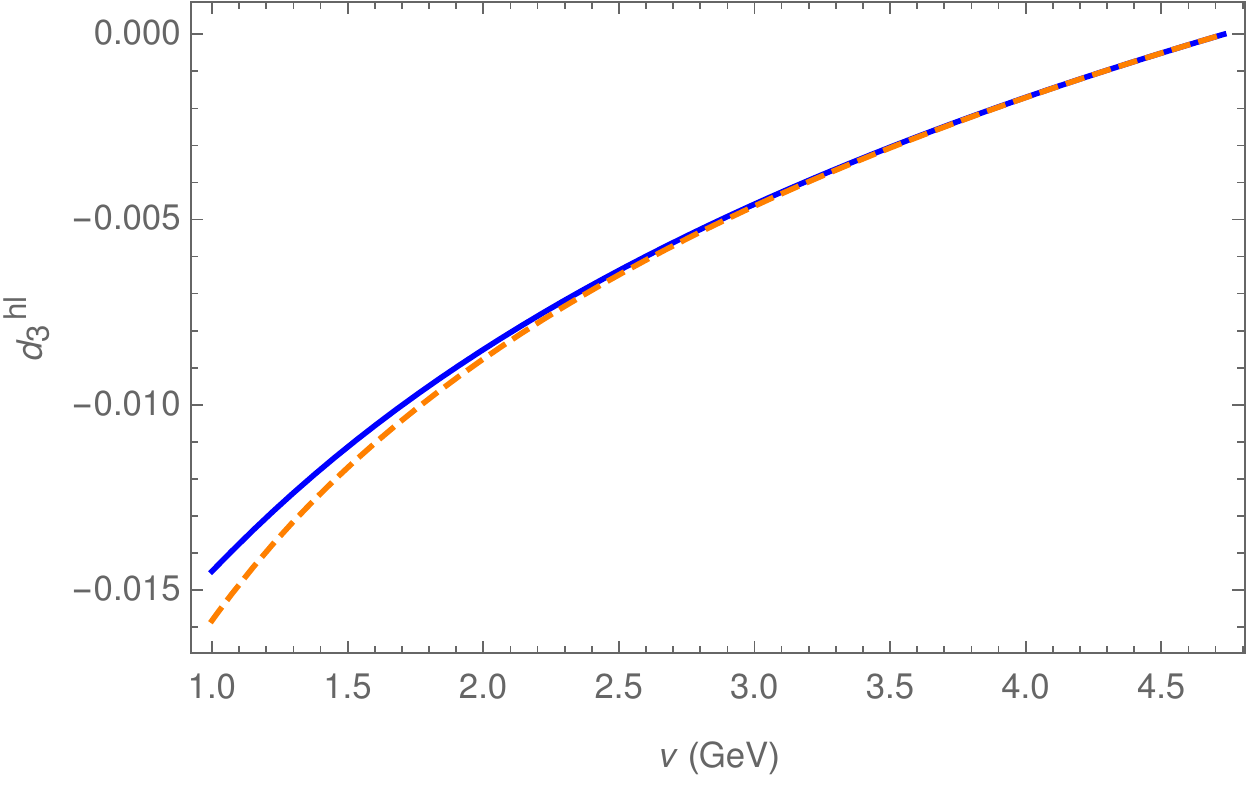}
	%
	\vspace{1ex}
	\includegraphics[width=0.43\textwidth]{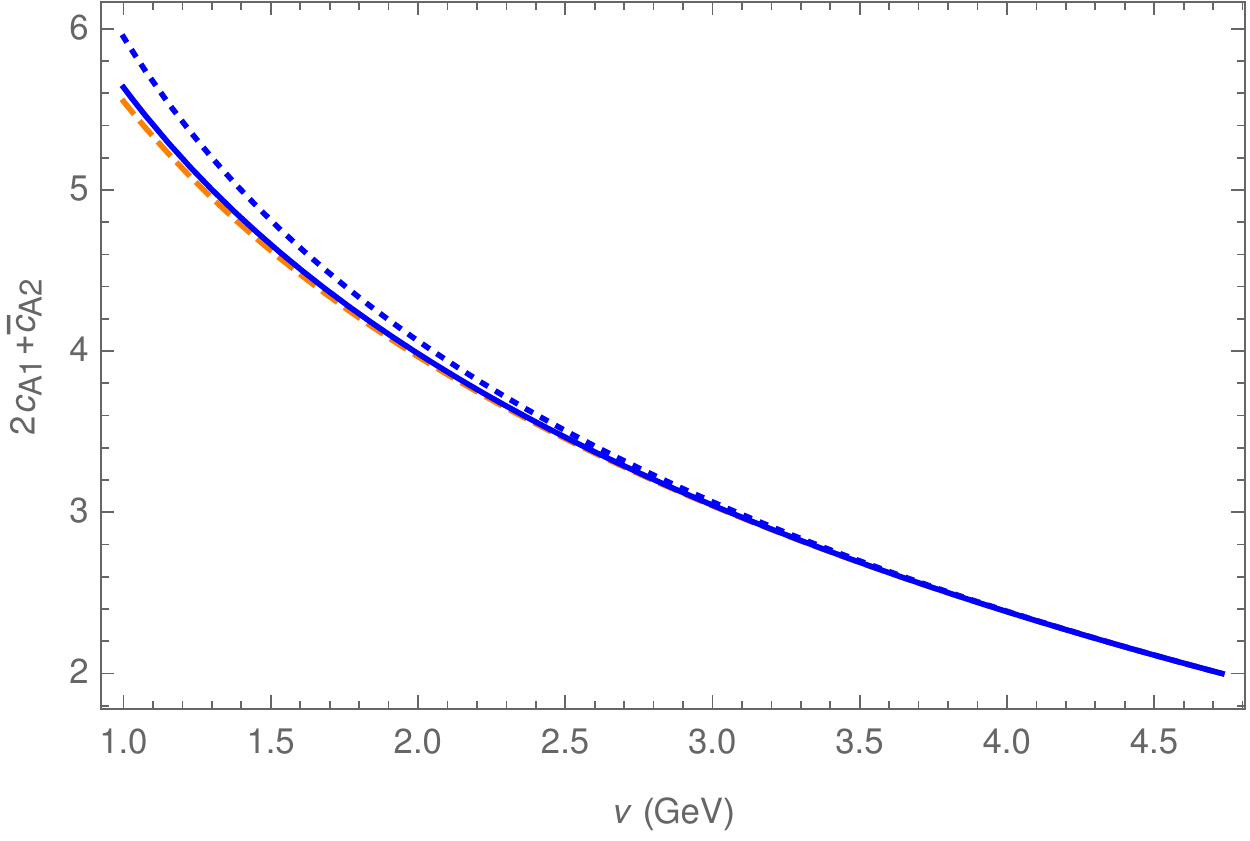}
\hspace{6ex}
	\includegraphics[width=0.43\textwidth]{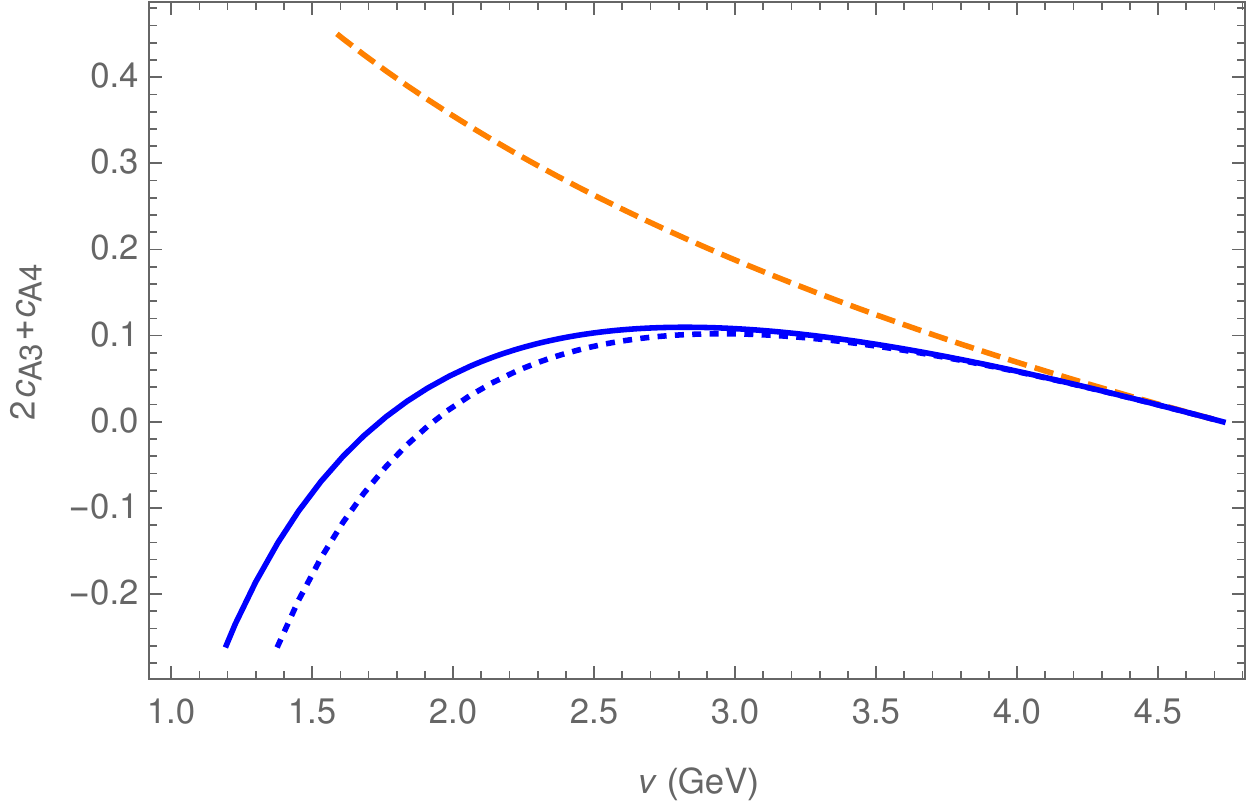}
	%
	%
\caption{Running of the $1/m^3$ spin-independent Wilson coefficients. The continuous line is the LL result with $n_f=4$, 
the dotted line is the LL result with $n_f=0$ and the dashed line is the single leading log result (it does not depend on $n_f$). 
\label{Plots}}   
\end{figure}

We draw in Fig. \ref{Plots} the above results when applied to the bottom heavy quark case to illustrate the importance of the incorporation of large logarithms in heavy quark physics (for physical processes where these Wilson coefficients appear). We run the Wilson coefficients from the heavy quark mass to 1 GeV for zero and four massless fermions. For illustrative purposes, we take $m_b=4.73$ GeV and $\al(m_b)=0.215943$. Besides the running of the Wilson coefficients, we also consider some specific combinations that appear in physical observables. In heavy quarkonium physics applications (like the NNNLL running of the spectrum and the NNLL running of Wilson coefficient of the electromagnetic current) we observe that only the combinations $\tilde c_{A_2} \equiv 2c_{A_1}+\bar c_{A_2}$,  $\tilde c_{A_4} \equiv 2c_{A_3}+ c_{A_4}$ appear \cite{inpreparation}. For the Compton scattering discussed in Sec.~\ref{Sec:Compton} we observe that $c_{A_{1,3}}$ and, again, $\tilde c_{A_{2,4}}$ appear. We remind the reader that those can be understood as linear combinations of the (chromo-)polarizabilities of the heavy quark.

For the numerical analysis we observe the following. For most cases the incorporation of light fermions plays a minor role in the result. The effect due to the logarithms are very large in most cases. This is due to very large coefficients multiplying the logs (even in the Abelian limit the coefficient is quite large). We also observe that the LL resummation is basically saturated by the single log in all cases except for  $c_{A_4}$ and $\tilde c_{A_4}$. Let us now discuss in more detail each individual Wilson coefficient. We observe the following: $c_{A_1}$ changes from 1 to -2 after running. The case of $\bar c_{A_2}$ is even more dramatic. It goes from 0 to 10 after running. The change after running of $c_{A_3}$ is more moderate though certainly sizable and so is for $c_{A_4}$. In this last case the resummation of logarithms happens to be important. Comparatively the running of $d_2^{hl}$ and $d_3^{hl}$ is much smaller, confirming that light fermions associated corrections are subleading. It is interesting to note that the running of $\tilde c_{A_2}$ is smaller than the running of $\bar c_{A_2}$ but still rather large, it changes by nearly a factor of 3. Finally, the qualitative behavior of $\tilde c_{A_4}$ and $ c_{A_4}$ is similar.

\subsection{Comparison with earlier work}
\label{Sec:Comparison}

The LL running of the Wilson coefficients of the $1/m^3$ operators of the  HQET Lagrangian was first addressed in Refs. \cite{Balzereit:1998jb,Balzereit:1998am,Balzereit:1998vh}. For the case with no light fermions, expressions for the anomalous dimension matrix and explicit expressions for the Wilson coefficients with single log accuracy are given. We find that these results are mutually inconsistent, as his anomalous dimension matrix produces a different expression for the explicit single log expression written in these references for the Wilson coefficients (except for $c_{12}^{(3)}$).

The basis of operators these results have been obtained is different from the basis we use in our paper. This makes the comparison difficult. In order to compare our results we have to move from one basis to the other. This is possible by the use of field redefinitions (at the order we are working it is equivalent to the use of the (full) equations of motion to order $1/m$). We obtain the 
following relation between the Wilson coefficients (spin-independent) in the two basis:
\bea
 c_1^{(3)}&=&-2c_M + \frac{1}{2}c_{A_2} + c_k c_F^2 + c_S c_F
\,,
\\
 c_2^{(3)} &=& c_4 + 2c_M -c_k^3 - c_D c_k
\,,
\\
 c_3^{(3)}&=&2c_M + c_{A_1} - c_k c_F^2 - c_S c_F
\,,
\\
 c_4^{(3)} &=& -4c_M - c_{A_1} + c_k c_F^2 + c_S c_F
\,,
\\
 c_{12}^{(3)} &=& \frac{1}{12}c_{A_3}
\,,
\\
c_{13}^{(3)} &=& \frac{1}{12}c_{A_4}
\,,
\eea
and these relations between the Wilson coefficients  at $\mathcal{O}(1/m)$ and $\mathcal{O}(1/m^2)$:
\bea
 c_1^{(1)} = c_k
\,,\qquad
 c_2^{(1)} = c_F
\,, \qquad
 c_1^{(2)} = -c_D
\,, \qquad
 c_2^{(2)} = c_S
\,.
\eea

In order to use these expressions we need to give values for $c_D$ and $c_M$. We use the LL resummed expressions of $c_D$ obtained in the Feynman gauge (for $c_M$ we use reparameterization invariance), as the computation in those references was done in the Feynman gauge. Note that this makes several of these coefficients gauge dependent. We can now produce expressions for the anomalous dimension matrix in Balzereit basis from our result. We obtain (we use the same ordering and notation as in Ref. \cite{Balzereit:1998jb}):
\begin{equation}
 \hat\gamma_l^{(3)A}=
 \begin{pmatrix}
   -11/12 & 0 & 11/24 & -11/24 & 11/288 & -11/72\\
   -4 & 0 & 1 & -1 & 0 & -2/9 \\
   -3 & 5/6 & 5/6 & - 5/3 & 1/12 & -7/18 \\
   -7/2 & 5/6 & 4/3 & -13/6 & 1/24 & -11/36 \\
   0 & 0 & 0 &0 &0 &0 \\
   0 & 0 & 0 &0 &0 &0 \\
   0 & 0 & 0 &0 &0 &0 \\
   0 & 0 & 0 &0 &0 &0 \\
   0 & 0 & 0 &0 &0 &0 \\
   0 & 0 & 0 &0 &0 &0 \\
   0 & 0 & 0 &0 &0 &0 \\
   0 & 0 & 0 &0 &0 &0 \\
   0 & 0 & 0 &0 &11/24 &-11/6 \\
 \end{pmatrix}
\end{equation}\\

\begin{equation}
 \hat\gamma_l^{(111)A}=
 \begin{pmatrix}
   16/3 & 0 & -5/2 & 7/3 & 0 & 0\\
   0 & 0 & 0 & 0 & 0 & 0 \\
   3 & -2/3 & -11/6 & 5/3 & -1/4 & 1/2 \\
   3/2 & 0 & -3/2 & 3/2 & 1/8 & -1/4 \\
 \end{pmatrix}
\end{equation}\\

\begin{equation}
 \hat\gamma_l^{(111)F}=
 \begin{pmatrix}
   32/3 & 0 & -8/3 & 0 & 0 & 0\\
   0 & 0 & 0 & 0 & 0 & 0 \\
   0 & 0 & 0 & 0 & 0 & 0 \\
   0 & 0 & 0 & 0 & 0 & 0 \\
 \end{pmatrix}
\end{equation}\\

\begin{equation}
 \hat\gamma_l^{(12)A}=
 \begin{pmatrix}
   4 & -5/6 & -3/2 & 2 & 0 & 2/9\\
   0 & 0 & 0 & 0 & 0 & 0 \\
   0 & 0 & 0 & 0 & 0 & 0 \\
   5/6 & 1/3 & -1/12 & -1/4 & 1/144 & 11/36 \\
 \end{pmatrix}
\end{equation}\\

\begin{equation}
 \hat\gamma_l^{(12)F}=
 \begin{pmatrix}
   0 & 0 & 0 & 0 & 0 & 0\\
   0 & 0 & 0 & 0 & 0 & 0 \\
   0 & 0 & 0 & 0 & 0 & 0 \\
    0 & 0 & 0 & 0 & 0 & 0\\
 \end{pmatrix}
\end{equation}\\
where we only include the contribution of spin-independent operators: $\mathcal{O}_{1-4,12,13}$. In all matrices (except the last one) we find discrepancies with the entries in the Appendix of Ref. \cite{Balzereit:1998jb}. The differences do not follow a clear pattern. On the other hand, remarkably enough, our anomalous dimension matrix produces the same single logs as those in Table II of Ref. \cite{Balzereit:1998vh} (note that the expression for $c_1^{(3)p}$ is different from the one one can find in Table I of Ref. \cite{Balzereit:1998jb}).

It is also interesting to make the comparison backward and try to produce results for our Wilson coefficients from the results obtained in Refs. \cite{Balzereit:1998jb,Balzereit:1998vh}. The inverse relations read:
\bea
 c_M &=& -\frac{1}{2}(c_3^{(3)} + c_4^{(3)})
\,,
\\
c_4 &=& c_2^{(3)} + c_3^{(3)} + c_4^{(3)} + c_1^{(1)\,3} - c_1^{(1)}c_1^{(2)}
\,,
\\
 c_{A_1} &=& 2c_3^{(3)} + c_4^{(3)} + c_1^{(1)}c_2^{(1)\,2} + c_2^{(1)}c_2^{(2)}
\,,
\\
 \bar c_{A_2} &=& 2c_1^{(3)} - 2c_1^{(1)}c_2^{(1)\,2} - 2c_2^{(2)}c_2^{(1)}
\,,
\\
 c_{A_3} &=& 12 c_{12}^{(3)}
\,,
\\
 c_{A_4} &=& 12 c_{13}^{(3)}
\,.
\eea
The running of $c_M$ agrees with the result predicted by reparametrization invariance (in Feynman gauge). For the gauge invariant combinations of 
Wilson coefficients we have computed in our paper, the anomalous dimension matrix given in Ref. \cite{Balzereit:1998jb} yields different RG equations than those we found in Sec. \ref{sec:RGeq} (nor even the running of $c_4$ is zero), and also different logs as those we found in  Eqs. (\ref{cA1SL}-\ref{cA4SL}) (except for (\ref{cA3SL})). On the other hand the single logs given in Table II of Ref. \cite{Balzereit:1998vh} yield results in agreement with our single log results in Eqs. (\ref{cA1SL}-\ref{cA4SL}).

The case including light fermions was analyzed in Ref. \cite{Balzereit:1998am}. In this case no anomalous dimension matrix was given but only the result for the single logs. For the operators we consider in our paper, such results disagree with Eqs. (\ref{d2hlSL}-\ref{d3hlSL}). 
We find $c_{3-}^{(3l)o}=8d_2^{hl}$ and $c_{3-}^{(3l)s}=8d_3^{hl}$, which is in disagreement by a factor of two.

\section{Conclusions}

We have computed the LL running of the Wilson coefficients of the spin-independent $1/m^3$ operators of the HQET Lagrangian in the case without light fermions. These include the heavy quark chromo-polarizabilities induced by the strong interactions. 
We have also computed the LL running of the Wilson coefficients of the spin-independent $1/m^3$ operators of the HQET Lagrangian in the case with light fermions except for the heavy-light operators that do not contribute to the chromo-polarizabilities. 

We have performed a numerical analysis of these results. We observe that the running produces a very large effect. For combinations that appear in physical cases such as in heavy quarkonium dynamics or in Compton scattering, the running is more moderate but still quite large.

These results are necessary building blocks for the complete determination of the production (and annihilation) of heavy quarkonium with NNLL accuracy near threshold and of the heavy quarkonium mass with NNNLL precision. 

\medskip

{\bf Acknowledgments} \\
We thank Matthias Steinhauser for the reading of the manuscript. 
This work was supported in part by the Spanish grants FPA2014-55613-P, 
FPA2013-43425-P and SEV-2016-0588.

\appendix

\section{HQET Feynman rules}

Here, we collect some Feynman rules (in Coulomb gauge) that we need for our computation, and complement those that can be found in \cite{Pineda:2011dg}. We also profit to correct a misprint of Eq. (205) of that reference in Eq. (\ref{FR:cS}). 

The heavy quark four-momentum $p=(E_1,\bf p)$ is incoming with associated color index $\beta$, 
whereas the heavy quark four-momentum $p'=(E_1',\bf p\,')$ is outgoing with associated color index $\alpha$. 

Incoming light quarks have four-momentum $-k_1$, color index $\gamma$ and gamma matrix index $A$, whereas outgoing light quarks have four-momentum $k_2$, color index $\delta$ and gamma matrix index $B$. Gluons appearing in Feynman rules with light quarks have outgoing four-momentum $k_3$ with color index $b$ and spatial vector index $i$ in case the gluon is transverse.

For gluons in Feynman rules with no light quarks the notation is the following. All four-momentums of gluons, $ k_i$, are outgoing. If more than one gluon appears, let's say $n$, 
then they are labeled with four-momentum $k_i$ ($i=1,\ldots,n$) and by four-momentum conservation $k=\sum_{i=1}^{n}k_i=p-p'$. We start labeling 
transverse gluons first and longitudinal gluons after with the labels $a,b,c,\ldots$ referring to color indices in the adjoint representation, $i, j, k,\ldots$ 
referring to space vector indices and $k_1, k_2, k_3,\ldots$ referring to four-momentum.

\subsection{Proportional to $c_i^{hl}$}

\begin{equation}
 \mathcal{V}= c_1^{hl}\frac{ig^2}{8m^2}(T^a)_{\alpha\beta}(T^a)_{\delta\gamma}(\gamma^0)_{BA}
\end{equation}

\begin{equation}
 \mathcal{V}= -c_2^{hl}\frac{ig^2}{8m^2}\bfsigma^i(T^a)_{\alpha\beta}(T^a)_{\delta\gamma}(\gamma^i \gamma_5)_{BA}
\end{equation}

\begin{equation}
 \mathcal{V}= c_3^{hl}\frac{ig^2}{8m^2}(I_{N_c})_{\alpha\beta}(I_{N_c})_{\delta\gamma}(\gamma^0)_{BA}
\end{equation}

\begin{equation}
 \mathcal{V}= -c_4^{hl}\frac{ig^2}{8m^2}\bfsigma^i(I_{N_c})_{\alpha\beta}(I_{N_c})_{\delta\gamma}(\gamma^i \gamma_5)_{BA}
\end{equation}

\subsection{Proportional to $c_D$}


\begin{equation}
 \mathcal{V}_{c_D}^{i\,ab}= c_D\frac{ig^2}{4m^2} {\bf k}_2^i[T^a,T^b]_{\alpha\beta}
\end{equation}

\begin{equation}
 \mathcal{V}_{c_D}^{ij\,ab}= c_D\frac{ig^2}{8m^2}\delta^{ij}(k_1^0-k_2^0)[T^a,T^b]_{\alpha\beta}
\end{equation}

\begin{equation}
 \mathcal{V}_{c_D}^{ij\,abc}= c_D\frac{ig^3}{8m^2}\delta^{ij}([T^a,[T^b,T^c]]_{\alpha\beta} + [T^b,[T^a,T^c]]_{\alpha\beta})
\end{equation}

\subsection{Proportional to $c_S$}

\begin{equation}
\label{FR:cS}
 \mathcal{V}_{c_S}^{a}= c_S\frac{g}{4m^2}{\bfsigma}\cdot({\bf p}\,'\times{\bf p})(T^a)_{\alpha\beta} 
\end{equation}

\begin{equation}
 \mathcal{V}_{c_S}^{i\,a}= -c_S\frac{g}{8m^2}k^0(\bfsigma\times({\bf p}+{\bf p}\,'))^i(T^a)_{\alpha\beta} 
\end{equation}

\begin{equation}
 \mathcal{V}_{c_S}^{i\,ab}= c_S\frac{g^2}{8m^2}[(\bfsigma\times({\bf p}+{\bf p}\,'))^i[T^a,T^b]_{\alpha\beta}
 +(\bfsigma_1\times {\bf k}_2)^i\{T^a,T^b\}_{\alpha\beta}]
\end{equation}

\begin{equation}
 \mathcal{V}_{c_S}^{ij\,ab}= c_S\frac{g^2}{8m^2}\bfsigma^k \epsilon^{kij}(k_2^0-k_1^0)\{T^a,T^b\}_{\alpha\beta}
\end{equation}

\begin{equation}
 \mathcal{V}_{c_S}^{ij\,abc}= -c_S\frac{g^3}{8m^2}\bfsigma^k \epsilon^{kij}
 (\{T^a,[T^b,T^c]\}_{\alpha\beta} - \{T^b,[T^a,T^c]\}_{\alpha\beta})
\end{equation}

\subsection{Proportional to $c_4$}

\begin{equation}
 \mathcal{V}_{c_4}^{i\,a}=-c_4\frac{ig}{8m^3}({\bf p}^2 +{\bf p}'^2)({\bf p}+{\bf p}')^i(T^a)_{\alpha\beta}
\end{equation}

\begin{equation}
 \mathcal{V}_{c_4}^{ij\,ab}=c_4\frac{ig^2}{8m^3}\left(\delta^{ij}({\bf p}^2+{\bf p}'^2)\{T^a,T^b\}_{\alpha\beta}
 +4{\bf p}'^i {\bf p}^j (T^a T^b)_{\alpha\beta} + 4{\bf p}^i {\bf p}'^j (T^b T^a)_{\alpha\beta}\right)
\end{equation}

\begin{eqnarray}\nonumber
&\mathcal{V}_{c_4}^{ijk\,abc}=& -c_4\frac{ig^3}{4m^3}\left[
\delta^{ij}\left( (T^c\{T^a,T^b\})_{\alpha\beta}{\bf p}'^k + (\{T^a,T^b\}T^c)_{\alpha\beta} {\bf p}^k \right)
\right.\\ && \left.
+\delta^{jk}\left( (T^a\{T^b,T^c\})_{\alpha\beta}{\bf p}'^i + (\{T^b,T^c\}T^a)_{\alpha\beta} {\bf p}^i\right)
\right.\\ && \left.
 + \delta^{ik}\left( (T^b\{T^a,T^c\})_{\alpha\beta}{\bf p}'^j + (\{T^a,T^c\}T^b)_{\alpha\beta} {\bf p}^j\right)
\right]
\end{eqnarray}

\subsection{Proportional to $c_M$}

\begin{equation}
 \mathcal{V}_{c_M}^{i\,a}=-c_M\frac{ig}{8m^3}{\bf k}^2({\bf p}+{\bf p}')^i(T^a)_{\alpha\beta}
\end{equation}

\begin{eqnarray}
\mathcal{V}_{c_M\,CG}^{ij\,ab}
&=&c_M\frac{ig^2}{8m^3}\left(\delta^{ij}\{T^a,T^b\}_{\alpha\beta}({\bf k}_1^2 +{\bf k}_2^2)
 - \delta^{ij}[T^a, T^b]_{\alpha\beta}( ({\bf p} + {\bf p}\,')\cdot({\bf k}_1 - {\bf k}_2) )
 \right)
\nn
\\
&&
\left. + 4[T^a,T^b]_{\alpha\beta}({\bf p}^i {\bf k}_1^j -{\bf k}_2^i {\bf p}^j)\right)
\eea
\begin{eqnarray}
\mathcal{V}_{c_M\,CG}^{ijk\,abc}&=& c_M\frac{ig^3}{8m^3}\left(
[T^a,[T^b,T^c]]_{\alpha\beta}(\delta^{ik}({\bf p}'+{\bf p})^j -\delta^{ij}({\bf p}'+{\bf p})^k)
\right.
\nn
\\
&&
+ [T^b,[T^a,T^c]]_{\alpha\beta}(\delta^{jk}({\bf p}'+{\bf p})^i -\delta^{ij}({\bf p}'+{\bf p})^k)
\nn
\\ && 
+ [T^c,[T^a,T^b]]_{\alpha\beta}(\delta^{jk}({\bf p}'+{\bf p})^i -\delta^{ik}({\bf p}'+{\bf p})^j)
\nn
\\
&&
+\{T^a,[T^b,T^c]\}_{\alpha\beta}(2\delta^{ik}{\bf k}_3^j -2\delta^{ij}{\bf k}_2^k +\delta^{jk}({\bf k}_2-{\bf k}_3)^i)
\nn
\\ 
&& 
+\{T^b,[T^a,T^c]\}_{\alpha\beta}(2\delta^{jk}{\bf k}_3^i -2\delta^{ij}{\bf k}_1^k +\delta^{ik}({\bf k}_1-{\bf k}_3)^j)
\nn
\\
&&
\left.
+\{T^c,[T^a,T^b]\}_{\alpha\beta}(2\delta^{jk}{\bf k}_2^i -2\delta^{ik}{\bf k}_1^j +\delta^{ij}({\bf k}_1-{\bf k}_2)^k)
\right)
\end{eqnarray}

\subsection{Proportional to $c_{A_1}$}

\begin{equation}
 \mathcal{V}_{c_{A_1}}^{ij\,ab}= c_{A_1}
\frac{ig^2}{8m^3}(\delta^{ij}(k_1^0 k_2^0 -{\bf k}_1\cdot{\bf k}_2) + {\bf k}_2^i {\bf k}_1^j)\{T^a,T^b\}_{\alpha\beta}
\end{equation}

\bea
 \mathcal{V}_{c_{A_1}}^{ijk\,abc}&=& c_{A_1}\frac{ig^3}{8m^3}
 \left[(\delta^{ij}{\bf k}_1^k -\delta^{ik}{\bf k}_1^j)\{[T^b,T^c],T^a\}_{\alpha\beta}
 + (\delta^{ij}{\bf k}_2^k -\delta^{jk}{\bf k}_2^i)\{[T^a,T^c],T^b\}_{\alpha\beta}
\right.
\nn
\\
&&
\left.
 +(\delta^{ik}k_3^j -\delta^{jk}{\bf k}_3^i)\{[T^a,T^b],T^c\}_{\alpha\beta}\right]
\eea

\begin{equation}
 \mathcal{V}_{c_{A_1}}^{i\,ab}= -c_{A_1}\frac{ig^2}{8m^3}k_1^0 {\bf k}_2^i\{T^a,T^b\}_{\alpha\beta}
\end{equation}

\begin{equation}
 \mathcal{V}_{c_{A_1}}^{ij\,abc}= -c_{A_1}\frac{ig^3}{8m^3}\delta^{ij}(k_1^0\{T^a,[T^b,T^c]\}_{\alpha\beta} +k_2^0\{T^b,[T^a,T^c]\}_{\alpha\beta})
\end{equation}

\begin{equation}
 \mathcal{V}_{c_{A_1}}^{ab}= c_{A_1}\frac{ig^2}{8m^3}{\bf k}_1\cdot{\bf k}_2\{T^a,T^b\}_{\alpha\beta}
\end{equation}

\begin{equation}
 \mathcal{V}_{c_{A_1}}^{i\, abc}= c_{A_1}\frac{ig^3}{8m^3}({\bf k}_2^i\{T^b,[T^a,T^c]\}_{\alpha\beta} + {\bf k}_3^i\{T^c,[T^a,T^b]\}_{\alpha\beta})
\end{equation}

\subsection{Proportional to $c_{A_2}$}

\begin{equation}
 \mathcal{V}_{c_{A_2}}^{ij\,ab}= c_{A_2}\frac{ig^2}{16m^3}\delta^{ij}k_1^0 k_2^0\{T^a,T^b\}_{\alpha\beta}
\end{equation}

\begin{equation}
 \mathcal{V}_{c_{A_2}}^{i\,ab}= -c_{A_2}\frac{ig^2}{16m^3}k_1^0 {\bf k}_2^i\{T^a,T^b\}_{\alpha\beta}
\end{equation}

\begin{equation}
 \mathcal{V}_{c_{A_2}}^{ij\,abc}= -c_{A_2}\frac{ig^3}{16m^3}\delta^{ij}(k_1^0\{T^a,[T^b,T^c]\}_{\alpha\beta} +k_2^0\{T^b,[T^a,T^c]\}_{\alpha\beta}])
\end{equation}

\begin{equation}
 \mathcal{V}_{c_{A_2}}^{ab}= c_{A_2}\frac{ig^2}{16m^3}{\bf k}_1\cdot{\bf k}_2\{T^a,T^b\}_{\alpha\beta}
\end{equation}

\begin{equation}
 \mathcal{V}_{c_{A_2}}^{i\, abc}= c_{A_2}\frac{ig^3}{16m^3}({\bf k}_2^i\{T^b,[T^a,T^c]\}_{\alpha\beta} + {\bf k}_3^i\{T^c,[T^a,T^b]\}_{\alpha\beta})
\end{equation}

\subsection{Proportional to $c_{A_3}$}

\begin{equation}
 \mathcal{V}_{c_{A_3}}^{ij\,ab}= c_{A_3}\frac{ig^2}{4m^3}\frac{T_F}{N_c}(\delta^{ij}(k_1^0 k_2^0 -{\bf k}_1\cdot{\bf k}_2) + {\bf k}_2^i {\bf k}_1^j)\delta^{ab}(I_{N_c})_{\alpha\beta}
\end{equation}

\begin{equation}
 \mathcal{V}_{c_{A_3}}^{ijk\,abc}= c_{A_3}\frac{g^3}{4m^3}\frac{T_F}{N_c} f^{abc}
 \left[\delta^{ik}({\bf k}_1-{\bf k}_3)^j +\delta^{jk}({\bf k}_3-{\bf k}_2)^i + \delta^{ij}({\bf k}_2-{\bf k}_1)^k\right](I_{N_c})_{\alpha\beta}
\end{equation}

\begin{equation}
 \mathcal{V}_{c_{A_3}}^{i\,ab}= -c_{A_3}\frac{ig^2}{4m^3}\frac{T_F}{N_c} k_1^0 {\bf k}_2^i\delta^{ab}(I_{N_c})_{\alpha\beta}
\end{equation}

\begin{equation}
 \mathcal{V}_{c_{A_3}}^{ij\,abc}= c_{A_3}\frac{g^3}{4m^3}\frac{T_F}{N_c}\delta^{ij}f^{abc}(k_1^0-k_2^0)(I_{N_c})_{\alpha\beta}
\end{equation}

\begin{equation}
 \mathcal{V}_{c_{A_3}}^{ab}= c_{A_3}\frac{ig^2}{4m^3}\frac{T_F}{N_c}{\bf k}_1\cdot{\bf k}_2 \delta^{ab}(I_{N_c})_{\alpha\beta}
\end{equation}

\begin{equation}
 \mathcal{V}_{c_{A_3}}^{i\, abc}= -c_{A_3}\frac{g^3}{4m^3}\frac{T_F}{N_c} f^{abc}({\bf k}_3-{\bf k}_2)^i(I_{N_c})_{\alpha\beta}
\end{equation}

\subsection{Proportional to $c_{A_4}$}

\begin{equation}
 \mathcal{V}_{c_{A_4}}^{ij\,ab}= c_{A_4}\frac{ig^2}{8m^3}\frac{T_F}{N_c}\delta^{ij}k_1^0 k_2^0\delta^{ab}(I_{N_c})_{\alpha\beta}
\end{equation}

\begin{equation}
 \mathcal{V}_{c_{A_4}}^{i\,ab}= -c_{A_4}\frac{ig^2}{8m^3}\frac{T_F}{N_c} k_1^0 {\bf k}_2^i\delta^{ab}(I_{N_c})_{\alpha\beta}
\end{equation}

\begin{equation}
 \mathcal{V}_{c_{A_4}}^{ij\,abc}= c_{A_4}\frac{g^3}{8m^3}\frac{T_F}{N_c}\delta^{ij}f^{abc}(k_1^0-k_2^0)(I_{N_c})_{\alpha\beta}
\end{equation}

\begin{equation}
 \mathcal{V}_{c_{A_4}}^{ab}= c_{A_4}^{(1)}\frac{ig^2}{8m^3}\frac{T_F}{N_c}{\bf k}_1\cdot{\bf k}_2 \delta^{ab}(I_{N_c})_{\alpha\beta}
\end{equation}

\begin{equation}
 \mathcal{V}_{c_{A_4}}^{i\, abc}= -c_{A_4}^{(1)}\frac{g^3}{8m^3}\frac{T_F}{N_c} f^{abc}({\bf k}_3-{\bf k}_2)^i(I_{N_c})_{\alpha\beta}
\end{equation}

\subsection{Proportional to $d_1^{hl}$}

\begin{equation}
 \mathcal{V}= -d_{1}^{hl}\frac{ig^2}{m^3}(T^a)_{\alpha\beta}(T^a)_{\delta\gamma}(\gamma^i)_{BA}({\bf p}+{\bf p}')^i
\end{equation}

\begin{equation}
 \mathcal{V}^{i\,b}= d_{1}^{hl}\frac{ig^3}{m^3}\{T^a, T^b\}_{\alpha\beta}(T^a)_{\delta\gamma}(\gamma^i)_{BA}
\end{equation}

\subsection{Proportional to $d_2^{hl}$}

\begin{equation}
 \mathcal{V}=-d_{2}^{hl}\frac{ig^2}{m^3}(T^a)_{\alpha\beta}(T^a)_{\delta\gamma}(\gamma^0)_{BA}(k_1^0-k_2^0)
\end{equation}

\begin{equation}
 \mathcal{V}^b= -d_{2}^{hl}\frac{ig^3}{m^3}(T^a)_{\alpha\beta}\{T^a,T^b\}_{\delta\gamma}(\gamma^0)_{BA}
\end{equation}

\subsection{Proportional to $d_3^{hl}$}

\begin{equation}
 \mathcal{V}=-d_{3}^{hl}\frac{ig^2}{m^3}(I_{N_c})_{\alpha\beta}(I_{N_c})_{\delta\gamma}(\gamma^0)_{BA}(k_1^0-k_2^0)
\end{equation}

\begin{equation}
 \mathcal{V}^b=- d_{3}^{hl}\frac{2ig^3}{m^3}(I_{N_c})_{\alpha\beta}(T^b)_{\delta\gamma}(\gamma^0)_{BA}
\end{equation}


\end{document}